\documentclass[left=2cm,%
right=2cm,%
top=2.25cm,%
bottom=2.25cm,%
headheight=11pt]{SelfArx} 

\usepackage[english]{babel} 
\usepackage{multirow,breqn,mathtools,makecell,caption}

\usepackage{lipsum} 
\graphicspath{ {figures/} }

\definecolor{color1}{RGB}{0,0,90} 
\definecolor{color2}{RGB}{0,20,20} 

\usepackage{hyperref} 
\hypersetup{hidelinks,colorlinks,breaklinks=true,urlcolor=color2,citecolor=color1,linkcolor=color1,bookmarksopen=false,pdftitle={Title},pdfauthor={Author}}

\JournalInfo{.} 
\Archive{.} 

\PaperTitle{A change of perspective in network centrality - Supporting Information\textsuperscript{+}} 

\Authors{Carla Sciarra\textsuperscript{1}*, Guido Chiarotti\textsuperscript{1},Francesco Laio\textsuperscript{1},Luca Ridolfi\textsuperscript{1}} 
\affiliation{\textsuperscript{1}\textit{Department of Environmental, Land and Infrastructure Engineering, Politecnico di Torino, Corso Duca degli Abruzzi, 24, 10129 Torino, (IT)}} 
\affiliation{*\textbf{Corresponding author}: carla.sciarra@polito.it} 
\affiliation{\textsuperscript{+} This is a pre-print of an article published in \textit{Scientific Reports}. The final authenticated version is available online at: \url{ https://doi.org/10.1038/s41598-018-33336-8}}

\newcommand{\keywordname}{Keywords} 

\Abstract{Typing ``Yesterday” into the search-bar of your browser provides a long list of websites with, in top places, a link to a video by The Beatles. The order your browser shows its search results is a notable example of the use of network centrality. Centrality is a measure of the importance of the nodes in a network and it plays a crucial role in a huge number of fields, ranging from sociology to engineering, and from biology to economics. Many metrics are available to evaluate centrality. However, centrality measures are generally based on \textit{ad hoc} assumptions, and there is no commonly accepted way to compare the effectiveness and reliability of different metrics. Here we propose a new perspective where centrality definition arises naturally from the most basic feature of a network, its adjacency matrix. Following this perspective, different centrality measures naturally emerge, including the degree, eigenvector, and hub-authority centrality. Within this theoretical framework, the accuracy of different metrics can be compared. Tests on a large set of networks show that the standard centrality metrics perform unsatisfactorily, highlighting intrinsic limitations of these metrics for describing the centrality of nodes in complex networks. More informative \textit{multi-component} centrality metrics are proposed as the natural extension of standard metrics.}

\begin{document}

\flushbottom 

\maketitle 


\thispagestyle{empty} 

In this supporting information, details about the mathematical results reported in the main text are provided. We start dealing with undirected networks, Sect. \ref{sect-1}, and then consider directed ones, Sect. \ref{sect-2}. 

\section{Undirected networks}
\label{sect-1}
\renewcommand{\thesubsection}{S1.\arabic{subsection}} 
\renewcommand{\theequation}{S1.\arabic{equation}}
\setcounter{equation}{0}
\subsection{General considerations}
In this work, we recast the problem of evaluating the centrality of the nodes in a network as a topology-estimation exercise. The estimator $\hat{A}_{ij}$ of the generic element $A_{ij}$ of the adjacency matrix depends on the centrality $x_i$ of the nodes, namely
\begin{equation}
\label{fund}
\hat{A}_{ij}= f(x_i, x_j).
\end{equation}
For undirected networks, the adjacency matrix $\mathbf{A}$ is symmetric, i.e., $A_{ij}=A_{ji}$. In our framework, this entails that the arguments of any estimator function $\hat{A}_{ij}$ should be exchangeable, namely 
\[
\hat{A}_{ij} = f (x_i, x_j) = f(x_j, x_i).
\]
The $x_i$ values are found by minimizing the sum of the squared ($S\! S$) residuals  between the original element $A_{ij}$ and its corresponding estimator $\hat{A}_{ij}$, with
\begin{equation}
	\label{SS}
	S\! S = \sum_i \sum_j (A_{ij}-\hat{A}_{ij})^2 = \sum_i \sum_j (A_{ij} - f(x_i,x_j))^2.
\end{equation}
The minimization procedure entails taking the derivative of $S\! S$ with respect to the considered variable (say, $x_k$), and equaling it to zero. $S\! S$ can be partitioned into two components: a first part which is independent of $x_k$ ($S\! S_0$), and a second part depending on $x_k$ ($S\! S_k$) i.e.,
\[
S\! S = S\! S_0 + S\! S_k.
\]
Notice that $S\! S_k$ only depends on the $k$-th row and column of the two matrices $\mathbf{A}$ and $\hat{\mathbf{A}}$, namely
\begin{align}
\label{ssk}
S\! S_k =& \sum_{i \neq k} \Bigl(A_{ik}- f(x_i,x_k) \Bigr)^2 + \sum_{j \neq k} \Bigl(A_{kj}- f(x_k,x_j) \Bigr)^2 \\ \nonumber & + \Bigl(A_{kk}- f(x_k,x_k) \Bigr)^2,
\end{align}
and the sums over the row and over the column coincide due to the symmetry of the matrix $\mathbf{A}$.

The derivative of the function $S\! S$ with respect to the variable $x_k$, using \eqref{ssk}, is
\begin{align}
\label{dssk1}
\frac{\partial S\! S}{\partial x_k} = \frac{\partial S\! S_k}{\partial x_k} =& 4 \sum_{i \neq k } \Bigr[ A_{ik}- f(x_i,x_k) \Bigr]~\frac{\partial f(x_i,x_k)}{\partial x_k} \\ \nonumber &+ 2 \Bigl[A_{kk}- f(x_k,x_k) \Bigr] ~\frac{\partial f(x_k,x_k)}{\partial x_k}=0. 
\end{align}
Noticing that
\[
\frac{\partial f(x_k,x_k)}{\partial x_k} = \frac{\partial f(x_i,x_k)}{\partial x_k}+\frac{\partial f(x_k,x_i)}{\partial x_k}= 2 \frac{\partial f(x_i,x_k)}{\partial x_k},
\]
\eqref{dssk1} becomes
\begin{equation}
\label{dss-und}
\frac{\partial S\! S_k}{\partial x_k} = 4 \sum_i \Bigr[ A_{ik}- f(x_i,x_k) \Bigr]~\frac{\partial f(x_i,x_k)}{\partial x_k}=0.
\end{equation}
From \eqref{dss-und}, $x_k$ is obtained. An equation equivalent to \eqref{dss-und} is obtained for any centrality value $x_i$, $(i=1,...,N)$.

Within the new perspective on network centrality described in this work, our definition of centrality is given through the analysis of the importance of the nodes in the estimation of $A_{ij}$, introducing the concept of unique contribution. We define the unique contribution of the generic node $k$ as the difference between the coefficient of determination describing the goodness of fit of the estimation $\hat{A}_{ij}$ considering all the $N$ centrality values, $R^2_{N}$, and the coefficient obtained by excluding the property of the node $k$, $R^2_{N-k}$. This yields 
\begin{equation}
\label{UC}
	UC_k=R^2_{N} - R^2_{N-k} = \frac{S\! S_{N-k} - S\! S_{N}}{T\! S\! S},
\end{equation}
in which we have used the definition
\[
R^2 = 1 - \frac{S\! S}{T\! S\! S},
\]
where $S\! S$ is defined in \eqref{SS}. $T\! S\! S$ is the variance of the adjacency matrix, i.e., $T\! S\! S= \sum_i \sum_j (A_{ij} - \bar{A})^2$, with $\bar{A}$ the mean of the matrix $\mathbf{A}$, namely
\[
\bar{A} = \frac{\sum_i \sum_j A_{ij}}{N^2} = \frac{K_{tot}}{N^2}.
\]
Hence 
\begin{align*}
T\! S\! S&=\sum_i \sum_j (A_{ij} - \bar{A}_{ij})^2 \\ \nonumber &= \sum_i \sum_j A_{ij}^2 - 2\frac{K_{tot}}{N^2}\sum_i \sum_j A_{ij} + \frac{K_{tot}^2}{N^2} \nonumber
\end{align*}
Since the elements of the adjacency matrix are either $1$ or $0$, $A_{ij}^2=A_{ij}$. This yields 
\begin{equation}
\label{TSS}
T\! S\! S = K_{tot} \Bigl( 1- \frac{K_{tot}}{N^2} \Bigr). 
\end{equation}

As obvious, $T\! S\! S$ does not change with the exclusion of $x_k$. In order to evaluate the unique contribution, it is hence sufficient to compute the variation $\Delta S\! S = S\! S_{N-k} - S\! S_{N}$ in \eqref{UC}. For the sake of simplicity, we are not repeating the estimation procedure without considering the variable $x_k$, but we are merely setting $x_k=0$ and keeping unchanged the other estimators $x_i$, $i \neq k$. Under these conditions, we can focus our attention on the $k$-th row and column only; $\Delta S\! S$ reads
\begin{align}
\label{deltass-gen}
\Delta S\! S =& 2 \sum_{i \neq k} \Bigl[ \Bigl( A_{ik} - f(x_i,0)\Bigr)^2 - \Bigl( A_{ik} - f(x_i,x_k)\Bigr)^2 \Bigr] \\ \nonumber 
&+ \Bigl( A_{kk} - f(0,0)\Bigr)^2 - \Bigl( A_{kk} - f(x_k,x_k)\Bigr)^2,
\end{align}
that can be expressed as
\begin{align}
\label{uc1}
\Delta S\! S =& 2 \sum_{i \neq k} \Bigl[ f(x_i,0)^2 - f(x_i,x_k)^2 -2f(x_i,0)A_{ik} +2 f(x_i,x_k)A_{ik} \Bigr] \\ \nonumber & + f(0,0)^2 - f(x_k,x_k)^2 -2f(0,0)A_{kk} + 2 f(x_k,x_k)A_{kk},
\end{align}
or 
\begin{align}
\label{uc2}
\Delta S\! S =& 2 \sum_{i \neq k} \Bigl( f(x_i,0)- f(x_i,x_k) \Bigr)\Bigl(f(x_i,0)+f(x_i,x_k)-2A_{ik} \Bigr) \\ \nonumber & + \Bigl( f(0,0) - f(x_k,x_k) \Bigr)\Bigl( f(0,0) + f(x_k,x_k) -2A_{kk} \Bigr).
\end{align}

Within this paper, we consider networks with no self-loops, hence $A_{kk}=0$.

\subsection{Degree centrality}
\label{sec-degree}
Let us start by considering the estimator $f_1$ for undirected networks,
\begin{equation}
\label{f1}
\hat{A}_{ij} = \mathit{f}_1(x_i,x_k) = a~\left[ x_i + x_k - \frac{1}{N}~\right].
\end{equation}

The derivative of the function $f_1$ with respect to $x_k$ is
\[
\frac{\partial f_1(x_i,x_k)}{\partial x_k} = a.
\]
Applying \eqref{dss-und} one obtains
\[
4 a \sum_i \Bigl[ A_{ik} - a \Bigl( x_i +x_k - \frac{1}{N}\Bigr) \Bigr] = 0.
\]
Since $\sum_i A_{ik} = k_k$ is the degree of the node $k$,  solving the equation for $x_k$ yields $x_k = \frac{k_k}{aN}$. Assuming the vector of centralities to have unitary 1-norm i.e., $\sum_i x_i =1$, one obtains
\begin{equation}\label{a}
a = \frac{K_{tot}}{N},
\end{equation}
finally yielding
\begin{equation}
\label{deg}
x_k = \frac{k_k}{K_{tot}}.
\end{equation}
\eqref{deg} corresponds to rescaling the \textbf{degree centrality} by the total degree of the network.

\subsubsection{Unique contribution}
From \eqref{f1}, one has
\[
f(x_i, 0) = a x_i - \frac{a}{N},
\]
and
\[
f(0,0)= - \frac{a}{N}.
\]
Using \eqref{uc2}, this provides
\begin{align*}
\Delta S\! S =& 2 \sum_{i \neq k} (-ax_k)\Bigl( 2ax_i + ax_k - 2\frac{a}{N} - 2 A_{ik} \Bigr) \\ \nonumber  & + (-2ax_k)\Bigl( 2ax_k -2\frac{a}{N} \Bigr) \\ \nonumber 
=& -2ax_k \sum_i \Bigl( 2ax_i +ax_k  - 2\frac{a}{N} + 2 A_{ik} \Bigr) + 2a^2x_k^2. \nonumber 
\end{align*}
Some further algebra provides
\[
\Delta S\! S = -2a^2x_k^2N + 4ax_kk_k+2a^2x_k^2.
\]
Substituting the value of $x_k$ as in \eqref{deg} and $a=K_{tot}/N$ in \eqref{a}, one obtains
\[
\Delta S\! S = \frac{2(N+1)k_k^2}{N^2}
\]
from which the unique contribution for the degree centrality is obtained,
\begin{equation}
\label{ucdeg}
UC_k = \frac{2(N+1)k_k^2}{N^2 T\! S\! S}.
\end{equation}
Since $UC_k$ is a monotonic increasing function of $k_k$, ranking for increasing $UC_k$ values provides the same ranking as the classical degree centrality.

\subsection{Eigenvector centrality}
\label{seig}
Consider the estimator for undirected network $f_2$ in Table 1, namely
\begin{equation}
\label{f2}
\hat{A}_{ik} = \mathit{f}_2(x_i,x_k) = \gamma x_i x_k.
\end{equation}
The derivative of the function $f_2$ with respect to $x_k$ is
\[
\frac{\partial f_2}{\partial x_k} = \gamma x_i
\]
Applying \eqref{dss-und} one obtains
\[
4 \sum_i \Bigl( A_{ik} -\gamma x_i x_k \Bigr)\gamma x_i = 0,
\]
that solved for $x_k$ provides
\[
x_k = \frac{\sum_i A_{ik}x_i}{\gamma \sum_i x_i^2}.
\]
We can assume the centrality vector to have unitary 2-norm (i.e., $\sum_i x_i^2 =1$). This yields
\begin{equation}
\label{eig}
x_k = \frac{1}{\gamma} \sum_i A_{ik}x_i.
\end{equation}
\eqref{eig} carries the same structure of the \textbf{eigenvector centrality} \cite{Newman}, where $\gamma=\lambda_1$ is the largest eigenvalue of $\mathbf{A}$. It is worth to notice that the relation in \eqref{SS}, with the function \eqref{f2}, recalls one of the relations from which Bonacich demonstrates the eigenvector centrality \cite{Bonacich1972}. However, this is just a formal resemblance; in fact, Bonacich used the Principal Factor Method, assuming $\mathbf{A}$ to be a special correlation matrix and $\mathbf{x}$ to be its first principal factor associated to the largest eigenvalue (see \cite{Bonacich1987,multivariate} for details).

\subsubsection{Unique contribution}
We use \eqref{uc1}, substituting $f_2$ for the generic function. In this case
\[
f(x_i,0)=f(0,0)=0,
\] 
from which \eqref{uc1} becomes
\begin{align*}
\Delta S\! S &= 2 \sum_{i \neq k} \Bigr[ -\gamma^2 x_i^2 x_k^2 + 2\gamma x_i x_k A_{ik} \Bigl] - \gamma^2 x_k^4 +2 \gamma x_k^2A_{kk} \\ \nonumber  
&= 2 \sum_i \Bigr[ -\gamma^2 x_i^2 x_k^2 + 2\gamma x_i x_k A_{ik} \Bigl] + \gamma^2 x_k^4 - 2 \gamma x_k^2A_{kk} \nonumber 
\end{align*}
Since the 2-norm of the vector is unitary, and using (see \eqref{eig}), 
\[
\sum_i A_{ik}x_i = \gamma x_k,
\]
one obtains
\[
\Delta S\! S = 2\gamma^2 x_k^2 + \gamma^2 x_k^4,
\]
in which the assumption $A_{kk}=0$ is used. Therefore, the unique contribution of the node, according to the definition in \eqref{UC}, is given by
\begin{equation}
\label{uceig}
UC_k = \frac{\gamma x_k^2}{T\! S\! S}\left(\gamma x_k^2+2\gamma\right).
\end{equation}
Since $UC_k$ is a monotonic increasing function of $x_k$, ranking for increasing $UC_k$ values provides the same ranking as the classical eigenvector centrality.

\subsection{Katz centrality}
\label{skatz}
Consider the estimation function $f_3$ for undirected networks (see Table 1) assuming the parameter $B$ to be negative, 
\begin{equation}
\label{f3}
\hat{A}_{ik} = \mathit{f}_3(x_i,x_k) = \gamma x_i x_k - \mathit{B}.
\end{equation}
The derivative of the function $f_3$ with respect to $x_k$, is
\[
\frac{\partial f_3}{\partial x_k} = \gamma x_i,
\]
from which the derivative of the function $S\! S$ according to \eqref{dss-und} is
\begin{align}
&4 \sum_i \Bigl(A_{ik} - \gamma x_i x_k + B \Bigr) \gamma x_i = \\ \nonumber  & \sum_i A_{ik} x_i -\gamma x_k \sum_i x_i^2 + B \sum_i x_i= 0,\nonumber 
\end{align}
that, solved for $x_k$, provides
\begin{equation} 
\label{katziter}
x_k = \frac{\sum_i A_{ik} x_i}{\gamma \sum_i x_i^2} + \frac{B \sum_i x_i}{\gamma \sum_i x_i^2}.
\end{equation}
We now introduce the \textit{attenuation factor} $\alpha$ of the Katz centrality \cite{katz1953} and define the equivalences 
\begin{equation}
\label{alphabeta}
\frac{1}{\gamma \sum_i x_i^2} = \alpha, \hspace{0.5cm} \frac{B  \sum_i x_i}{\gamma \sum_i x_i^2} = \beta
\end{equation}
obtaining
\begin{equation}
\label{katz}
x_k = \alpha \sum_i A_{ik}x_i + \beta.
\end{equation}
\eqref{katz} corresponds to the definition of the Katz centrality measure \cite{katz1953}, in which $\alpha$ is the attenuation factor whose value is $\alpha<1/\lambda_1$, being $\lambda_1$ the largest eigenvalue of $\mathbf{A}$ and $\beta$ is a constant, whose value is usually set to one \cite{Newman}. Due to the constraint imposed by the form of the Katz centrality, the $x_i$ values are always positive and greater than one; hence no assumptions can be made on the norms of the vector $\mathbf{x}=[x_1,...,x_N]$. 

\subsubsection{Unique contribution}
Using the function $f_3$ in \eqref{f3}, one has
\[
f(x_i,0)=f(0,0)=-B.
\]
Using the form of $\Delta S\! S$ as given in \eqref{uc2} and substituting the values of the functions 
\[
f(x_i,0) - f(x_i,x_k) = -\gamma x_i x_k,
\hspace{0.5cm}
f(0,0) - f(x_k,x_k) = -\gamma x_k^2,
\]
one obtains 
\begin{align*}
\Delta S\! S = &2 \sum_{i \neq k} (-\gamma x_i x_k) (\gamma x_i x_k -2B -2A_{ik}) \\ \nonumber & - \gamma x_k^2 (\gamma x_k^2 -2B - 2A_{kk}) \\ \nonumber  = & 2 \sum_i (-\gamma^2 x_i^2 x_k^2 +2 \gamma B x_i x_k +2\gamma x_i x_k A_{ik}) \\ \nonumber & + \gamma^2 x_k^4 - 2\gamma B x_k^2 \nonumber,
\end{align*}
where the assumption $A_{kk}=0$ is used. Using the equivalences in \eqref{alphabeta}, and the one deriving from \eqref{katz}, 
\[
\sum_i A_{ik}x_i = \frac{x_k}{\alpha} - \frac{\beta}{\alpha},
\]
one obtains
\begin{align*}
\Delta S\! S =& -2 \gamma^2 x_k^2 \frac{1}{\alpha \gamma}+ 4 \gamma B x_k \frac{\beta}{\alpha B }+4\gamma x_k \Bigl(\frac{x_k}{\alpha} - \frac{\beta}{\alpha} \Bigr) \\ \nonumber & +\gamma^2 x_k^4 -2B\gamma x_k^2 \\ \nonumber = &~2 \gamma \frac{x_k^2}{\alpha} + \gamma^2 x_k^4 -2B\gamma x_k^2. \nonumber 
\end{align*}
The unique contribution of the node, according to the definition \eqref{UC} is given by
\begin{equation}
\label{uckatz}
UC_k = \frac{\gamma x_k^2}{T\! S\! S}\left(\gamma x_k^2-2B+ \frac{2}{\alpha} \right).
\end{equation}
Since we have defined $B$ to be negative, while $\gamma$ and $\alpha$ are positive, $UC_k$ is a monotonic increasing function of $x_k$ and ranking for increasing $UC_k$ values provides the same ranking as the classical Katz centrality. 
 
\subsection{Multi-component centrality}
\label{sec-mcu}
Within our change of perspective, we introduced multi-component centrality metrics to improve the quality of the estimation. Within this framework, in case of undirected network, the multidimensional estimator reads 
\begin{align}
\label{f_sequence}
\hat{A}_{ij}(s) &= \gamma_1 x_{i,1}x_{j,1} +\gamma_2 x_{i,2}x_{j,2}+...+ \gamma_s x_{i,s}x_{j,s} \\ \nonumber 
& = \sum_{t=1}^s \gamma_t x_{i,t} x_{j,t}.
\end{align}
The estimator is a function of the $s$-dimensional vector embedding the $s$ properties of the node that are considered for evaluating node's importance, namely $\hat{A}_{ij} = f(\mathbf{x}_i,\mathbf{x}_j)$, where $\mathbf{x}_i= [ x_{i,1},...,x_{i,s} ]$.

We assume the 2-norm of each vector $\mathbf{x}_t=[x_{1,t},...,x_{N,t}]$ to be unitary, i.e. $\sum_i x_{i,t}^2 =1$. Moreover, we set an orthogonality condition between any two vectors $\mathbf{x}_t$ and $\mathbf{x}_t^*$, i.e.
\begin{equation}
\label{ortho}
\sum_i x_{i,t} \cdot x_{i,t^{*}} =0, \hspace{0.5cm} \forall t \neq t^{*}.
\end{equation}
The steps described for the one-component centrality can be adapted to the multidimensional setting. 
In this setting, we consider the contribution to $S\! S$ of a generic variable $x_{k,t^{*}}$. As before, $S\! S$ is partitioned into a part $S\! S_0$, which does not depend on $x_{k,t^*}$, and a part $S\! S_{k,t^*}$, which is a function of $x_{k,t^*}$, 
\begin{equation}
\label{sskt}
S\! S = S\! S_{0,t} + S\! S_{k,t^{*}}.
\end{equation}
The computation of the centrality values by minimization of the $S\! S_{k,t*}$, entails computing \eqref{dss-und} accounting for each dimension considered i.e., $t={1,...,s}$. The derivative of $S\! S$ has the same form as \eqref{dss-und}. Using
\[
\frac{\partial f(\mathbf{x}_i,\mathbf{x}_k)}{\partial x_{k,t*}} = \gamma_{t*} x_{i,t*}.
\]
one obtains
\[
4 \sum_i \Bigl[ A_{ik} - \sum_t \gamma_t x_{i,t} x_{k,t} \Bigr] \gamma_{t^*} x_{i,t^*}=0,
\]
that is equivalent to
\[
\sum_i A_{ik}x_{i,t^*} - \sum_t \gamma_{t} x_{k,t} \sum_i x_{i,t}\cdot x_{i,t^*}=0.
\]
Due to the orthonormality condition set in \eqref{ortho}, it holds
\begin{align*}
\sum_t \gamma_{t} x_{k,t} \sum_i x_{i,t} \cdot x_{i,t^*} & =  \gamma_{t^*} x_{k,t^*} \sum_i x_{i,t^*} \cdot x_{i,t^*} \\ \nonumber &= \gamma_{t^*} x_{k,t^*} \sum_i x_{i,t^*}^2 = \gamma_{t^*} x_{k,t^*} \nonumber.
\end{align*}
Finally, for any component $t$, the centrality value reads
\begin{equation}
\label{eigt}
x_{k,t} = \frac{1}{\gamma_t} \sum_i A_{ik} x_{i,t},
\end{equation}
which corresponds to computing the eigenvector $\mathbf{x}_t$ corresponding to the eigenvalue $\gamma_t$.

In \eqref{f_sequence}, the eigenvalues $\gamma_t$, and hence their corresponding eigenvectors $\mathbf{x}_t$, can be ordered according to their absolute value. This solution corresponds to the \textit{Singular Value Decomposition} for symmetric matrices \cite{golub2012}, being $\hat{\mathbf{A}}(s)$ the \textit{s-order low-rank approximation} of the original adjacency matrix $\mathbf{A}$. The \textit{Eckhart-Young-Mirsky theorem} \cite{eckart1936} proofs that the total amount of explained variance $SSE$ of the \textit{s-order low-rank approximation} equals the sum of the squares of the $s$ eigenvalues, when the approximation is truncated at $s$, namely
\begin{equation}
\label{exp-var}
SSE(s) = \sum_{t=1}^s \gamma_t^2.
\end{equation}
For choosing the value of $s$, different strategies can be pursued (see, e.g., \cite{skillicorn2007} for a review of the criteria).
For a given number of components $s$, at each component $t^*$ added to the estimation, the total amount of explained variance increases by $\gamma_{t^*}^2$. Hence it holds
\begin{equation}
\label{sse-s-1}
SSE(t^*)-SSE(t^*-1) = \gamma_{t^*}^2
\end{equation}
The total amount of unexplained variance $SSU$ is
\begin{align}
\label{ssu}
SSU(t^*) &= \sum_i \sum_j \Bigl(A_{ij} - \hat{A}_{ij}(t^*)\Bigr)^2 = T\! S\! S - SSE(t^*) \\ \nonumber &= T\! S\! S - \sum_{t=1}^{t*} \gamma_{t^*}^2,
\end{align}
with $T\! S\! S$ as in \eqref{TSS}.

The ordering of the eigenvalues, however, requires some additional considerations. In fact, \eqref{exp-var} ensures that the explained variance with $s$ components is maximized by taking the first $s$ eigenvalues, ordered in absolute values from the largest to the smallest. However, a consistency issue emerges when considering networks with no self loops. For these networks, the elements on the diagonal of $\mathbf{A}$ are zero. The estimated matrix has instead its diagonal elements different from zero, namely
\begin{equation}
\label{trace}
\hat{A}_{ii}(s)= \sum_{t=1}^{s} \gamma_t x_{i,t}^2.
\end{equation}
This entails that, in order to provide a good description of the system, the eigenvalues should be ordered according to the total amount of explained variance they bring off-diagonal. In fact, \eqref{exp-var} can be partitioned in two terms, one pertaining with the diagonal $D$ and the other with the off-diagonal $OD$ terms, i.e., 
\begin{equation}
SSE(s) = SSE(s)_{D} + SSE(s)_{OD}.
\label{sse}
\end{equation}
We are therefore interested in ordering the eigenvalues so that the value $SSE(t^*)_{OD}$ at each new added component $t^*$ is maximized.

Consider the term $SSE(t^*)_D$. Using \eqref{ssu} and \eqref{trace}, this term reads
\begin{align}
\label{sse-s-2}
SSE(t*)_D &= T\! S\! S - \sum_i  \Bigl(A_{ii} - \hat{A}_{ii}(t^*)\Bigr)^2 = T\! S\! S - \sum_i \Bigl(\hat{A}_{ii}(t^*) \Bigr)^2 \\ \nonumber & = T\! S\! S - \sum_i\Bigl( \sum_{t=1}^{t^*} \gamma_t x_{i,t}^2\Bigr)^2
\\ \nonumber & = T\! S\! S - \sum_i \Bigl( \sum_{t=1}^{t^*-1} \gamma_t x_{i,t}^2 + \gamma_{t^*} x_{i,t^*}^2\Bigr)^2 \\ \nonumber & = T\! S\! S -\sum_i \Bigl(\sum_{t=1}^{t^*-1} \gamma_t x_{i,t}^2 \Bigr)^2 - \gamma_{t^*}^2 \sum_i x_{i,t^*}^2 \\ \nonumber  &- \sum_i 2 \gamma_{t^*} x_{i,t^*}^2 \Bigl( \sum_{t=1}^{t^*-1} \gamma_t x_{i,t}^2 \Bigr).
\end{align}
\eqref{sse-s-2} entails that at each new component $t=t^*$ added to the estimation, the increment in the total amount of explained variance on the diagonal $\Delta SSE(t^*)_D=SSE(t^*)_D - SSE(t^*-1)_D$ equals
\begin{equation}
\label{delta-sse-d}
\Delta SSE(t^*)_D = -\gamma_{t^*}^2 \sum_i x_{i,t^*}^4 - 2 \sum_i \gamma_{t^*} x_{i,t^*}^2 \sum_{t=1}^{t^*-1} \gamma_{t} x_{i,t}^2.
\end{equation}
Considering that the total amount of explained variance by the $t^*$ component is $\gamma_{t^*}^2$ (see \eqref{sse-s-1}), one obtains from \eqref{sse} and \eqref{delta-sse-d}
\begin{equation}
\label{delta-sse-od}
\Delta SSE(t^*)_{OD}= \gamma_{t^*}^2 \Bigl(1+ \sum_i x_{i,t^*}^4 \Bigr) + 2 \sum_i \gamma_{t^*} x_{i,t^*} \sum_{t=1}^{t^*-1} \gamma_t x_{i,t}^2.
\end{equation}
Aiming at choosing the order in which the eigenvalues, and respective eigenvectors, should be embedded into the estimation \eqref{f_sequence}, one should maximize, at each step, the function in \eqref{delta-sse-od}. For $t=1$ -- i.e., for choosing the first eigenvalue and respective eigenvector -- the function to be maximized is
\[
\Delta SSE(t^*=1)_{OD} = \gamma_{t}^2 \Bigl(1+ \sum_i x_{i,t}^4 \Bigr).
\]
When $t=2$, the second eigenvalue to be embedded into the function \eqref{f_sequence} is the one that maximizes the function
\[
\Delta SSE(t^*=2)_{OD} = \gamma_{t=1}^2 \Bigl(1+ \sum_i x_{i,t=1}^4 \Bigr) + 2 \gamma_{t=1} \gamma_t \sum_i x_{i,t}^2 x_{i,t=1}^2.
\]

In the main text, all of the results shown referring to the multi-component estimator and centrality have been processed according to the just described algorithm. 

\subsubsection{Unique contribution}
In the multi-component setting, the unique contribution is found accounting for all components $\mathbf{x}_t$, $t=(1,...,s)$. In this case, excluding the generic node $k$ from the estimation corresponds to nullifying all of its properties $x_{k,t}$, with $t=(1,...,s)$. This yields
\[
f(\mathbf{x}_i,0)=f(0,0)=0.
\]
Within this multi-component setting, \eqref{uc2} becomes
\begin{align*}
\Delta S\! S = &~ 2 \sum_{i \neq k} \Bigl( - \sum_{t=1}^s \gamma_t x_{i,t} x_{k,t} \Bigr) \Bigl( \sum_{t=1}^s \gamma_t x_{i,t} x_{k,t} - 2A_{ik} \Bigl)\\ \nonumber  &- \Bigl( \sum_{t=1}^s \gamma_t x_{k,t}^2 \Bigr) \Bigl( \sum_{t=1}^s \gamma_t x_{k,t}^2 -2A_{kk} \Bigr)
\\ \nonumber =&~ 2 \sum_i \Bigl[ - \Bigl( \sum_{t=1}^s \gamma_t x_{i,t} x_{k,t} \Bigr)^2 + 2A_{ik} \sum_{t=1}^s \gamma_t x_{i,t} x_{k,t}\Bigl]
\\ \nonumber &+ \Bigl( \sum_{t=1}^s \gamma_t x_{k,t}^2 \Bigr)^2. \nonumber 
\end{align*}
that is equivalent to
\begin{align*}
\Delta S\! S =& -2 \sum_{t=1}^s \gamma_t^2 x_{k,t}^2 \sum_i x_{i,t}^2 + 4 \sum_{t=1}^s \gamma_x x_{k,t} \sum_i A_{ik}x_{i,t} \\ \nonumber  &+ \Bigl( \sum_{t=1}^s \gamma_t x_{k,t}^2 \Bigr)^2 \nonumber 
\end{align*}
Using the orthonormality condition \eqref{ortho} and \eqref{eigt}, the unique contribution in the case of the multi-component estimator is given by
\begin{equation}
\label{uc-mult}
UC(s)_k = 2 \sum_{t=1}^s \gamma_t^2 x_{k,t}^2 + \Bigl( \sum_{t=1}^s \gamma_t x_{k,t}^2 \Bigr)^2.
\end{equation}

\section{Directed Networks}
\label{sect-2}
\renewcommand{\thesubsection}{S2.\arabic{subsection}} 
\renewcommand{\theequation}{S2.\arabic{equation}}
\setcounter{equation}{0}
\subsection{General considerations}
Consider a directed network, whose adjacency matrix $\mathbf{A}$ is generally asymmetric. The estimator $\hat{A}_{ij}$ of the generic element $A_{ij}$ now depends on both the \textit{out} and \textit{in} centrality of the nodes, namely
\begin{equation}
\label{fdir}
\hat{A}_{ij} = f(x_i^{out},x_j^{in}).
\end{equation}

The steps described for undirected networks to obtain the centrality values and to compute the unique contribution, Sect. \ref{sect-1}, can be easily adapted to directed networks. The minimization of the function $S\! S_k$ here corresponds to deriving the function with respect to the considered variables, $x_k^{out}$ and $x_k^{in}$, accounting for the asymmetry of $\mathbf{A}$. Hence, $S\! S_k$ reads
\begin{align}
\label{ssk-dir}
S\! S_k =& \sum_{i \neq k} \Bigl(A_{ik}- f(x_i^{out},x_k^{in}) \Bigr)^2 + \sum_{j \neq k} \Bigl(A_{kj}- f(x_k^{out},x_j^{in}) \Bigr)^2 \\ \nonumber  & + \Bigl(A_{kk}- f(x_k^{out},x_k^{in}) \Bigr)^2, 
\end{align}
In the case of directed networks, the arguments of the function are exchangeable only on the diagonal, namely
\[
f(x_k^{out},x_k^{in})=f(x_k^{in},x_k^{out}).
\]
The derivatives of the function $S\! S$ with respect to the variables $x_k^{out}$ and $x_k^{in}$ are
\begin{align}
\label{eq3-2}
\frac{\partial S\! S_k}{\partial x_k^{out}} =& ~ 2 \sum_{j \neq k} \Bigr[ A_{kj} - f(x_k^{out},x_j^{in}) \Bigl] \frac{\partial f(x_k^{out},x_j^{in})}{\partial x_k^{out}} \\ \nonumber &+ 2 \Bigl[A_{kk} -f(x_k^{out},x_k^{in})\Bigr] \frac{\partial f(x_k^{out},x_k^{in})}{\partial x_k^{out}} =0 ,
\end{align}
and
\begin{align}
\label{eq3-3}
\frac{\partial S\! S_k}{\partial x_k^{in}} =& ~ 2 \sum_{i \neq k} \Bigr[ A_{ik} - f(x_i^{out},x_k^{in}) \Bigl] \frac{\partial f(x_i^{out},x_k^{in})}{\partial x_k^{in}} \\ \nonumber &+ 2 \Bigl[A_{kk} -f(x_k^{out},x_k^{in})\Bigr] \frac{\partial f(x_k^{out},x_k^{in})}{\partial x_k^{in}} =0.
\end{align}
In \eqref{eq3-2} and \eqref{eq3-3}, both the terms $i=k$ and $j=k$ can be included into the sums. Hence
\begin{equation}
\label{dssout}
\frac{\partial S\! S_k}{\partial x_k^{out}} = 2 \sum_j \Bigr[ A_{kj} - f(x_k^{out},x_j^{in}) \Bigl] \frac{\partial f(x_k^{out},x_j^{in})}{\partial x_k^{out}} =0,
\end{equation}
and
\begin{equation}
\label{dssin}
\frac{\partial S\! S_k}{\partial x_k^{in}} = 2 \sum_i \Bigr[ A_{ik} - f(x_i^{out},x_k^{in}) \Bigl] \frac{\partial f(x_i^{out},x_k^{in})}{\partial x_k^{in}}=0.
\end{equation}

The unique contribution is found through \eqref{UC}, hence computing $\Delta S\! S = S\! S_{N-k} - S\! S_{N}$. In directed networks, nodes are characterized by two properties. Within this framework, the unique contribution can be computed with respect to one of the properties, or at the need, with respect to both ones. In the first case, one finds the \textit{in}-centrality (or the \textit{out}-centrality) of the node. In the second case the overall centrality of the node is obtained.

If both properties are considered in the computation, we can define $\Delta S\! S$ as
\begin{align}
\label{dss-dir}
\Delta S\! S^{tot} = & \sum_{i \neq k} \Bigl[ \Bigl( A_{ik} - f(x_i^{out},0)\Bigr)^2 - \Bigl( A_{ik} - f(x_i^{out},x_k^{in})\Bigr)^2 \Bigr] \\ \nonumber & \sum_{j \neq k} \Bigl[ \Bigl( A_{kj} - f(0,x_j^{in})\Bigr)^2 - \Bigl( A_{ik} - f(x_k^{out},x_j^{in})\Bigr)^2 \Bigr]\\ &+ \Bigl( A_{kk} - f(0,0)\Bigr)^2 - \Bigl( A_{kk} - f(x_k^{out},x_k^{in})\Bigr)^2,\nonumber 
\end{align}
in which we consider the exclusion of the properties $x_k^{out}$ and $x_k^{in}$ to be equivalent to setting $x_k^{out}=x_k^{in}=0$. \eqref{dss-dir} can be expressed as
\begin{align}
\label{uc-dir1}
\Delta S\! S^{tot} =& \sum_{i \neq k} \Bigl[ f(x_i^{out},0)^2 - f(x_i^{out},x_k^{in})^2 -2f(x_i^{out},0)A_{ik} \\ \nonumber & +2 f(x_i^{out},x_k^{in})A_{ik} \Bigr]+ \sum_{j \neq k} \Bigl[ f(0, x_j^{in})^2 - f(x_i^{out},x_k^{in})^2 \\ & -2f(0,x_j^{in})A_{kj} +2 f(x_i^{out},x_k^{in})A_{kj} \Bigr] + f(0,0)^2 \\ \nonumber & - f(x_k^{out},x_k^{in})^2 -2f(0,0)A_{kk} + 2 f(x_k^{out},x_k^{in})A_{kk}, 
\end{align}
or 
\begin{align}
\label{uc-dir2}
\Delta S\! S^{tot} =& \sum_{i \neq k} \Bigl( f(x_i^{out},0)- f(x_i^{out},x_k^{in}) \Bigr)\Bigl(f(x_i^{out},0)+f(x_i^{out},x_k^{in})\\ \nonumber &-2A_{ik} \Bigr) + \sum_{j \neq k} \Bigl( f(0,x_j^{in},0)- f(x_k^{out},x_k^{in}) \Bigr)\Bigl(f(0,x_j^{in}) \\ \nonumber &+f(x_k^{out},x_k^{in})-2A_{kj} \Bigr) + \Bigl( f(0,0) - f(x_k^{out},x_k^{in}) \Bigr)\\ \nonumber & \cdot\Bigl( f(0,0) - f(x_k^{out},x_k^{in}) -2A_{kk} \Bigr). 
\end{align}
The unique contribution is then found deploying the expression in \eqref{uc-dir1} or \eqref{uc-dir2}, and applying the definition in \eqref{UC}.

To compute the unique contribution with respect to one of the two properties entails considering, in \eqref{uc-dir1} or \eqref{uc-dir2}, only the terms on the dimension related to the specific property at hand. Hence, the $k$-th row (sum over $j$) for the \textit{out} centrality of the node $k$ and the $k$-th column (sum over $i$) for its \textit{in} centrality. In formulas
\begin{align}
\label{dss-out}
\Delta S\! S^{out} =& \sum_j \Bigl[ f(0, x_j^{in})^2 - f(x_k^{out},x_j^{in})^2 -2f(0,x_j^{in})A_{kj} \\ \nonumber &+2 f(x_k^{out},x_j^{in})A_{kj} \Bigr] \\ \nonumber =& \sum_j \Bigl( f(0,x_j^{in})- f(x_k^{out},x_j^{in}) \Bigr)\Bigl(f(0,x_j^{in}) +f(x_k^{out},x_j^{in}) \\ \nonumber & -2A_{kj} \Bigr), 
\end{align}
and
\begin{align}
\label{dss-in}
\Delta S\! S^{in} =& \sum_i \Bigl[ f(x_i^{out},0)^2 - f(x_i^{out},x_k^{in})^2 -2f(x_i^{out},0)A_{ik} \\ \nonumber & +2 f(x_i^{out},x_k^{in})A_{ik} \Bigr] \\ \nonumber =& \sum_i \Bigl( f(x_i^{out},0)- f(x_i^{out},x_k^{in}) \Bigr)\Bigl(f(x_i^{out},0)+f(x_i^{out},x_k^{in})\\ \nonumber &-2A_{ik} \Bigr),
\end{align}

In the following, we consider networks with no self-loops, hence $A_{kk}=0$.

\subsection{Degree centrality}
\label{sec-deg-dir}
Consider the function $f_1$
\begin{equation}
\label{f1dir}
\hat{A}_{ij} = f_1 (x_i^{out},x_k^{in}) = a \Bigl[ x_i^{out} + x_k^{in} - \frac{1}{N} \Bigr].
\end{equation}
The derivatives of the function $f_1$ with respect to both properties $x_k^{out}$ and $x_k^{in}$ are
\[
\frac{\partial f_1}{\partial x_k^{out}} = \frac{\partial f_1}{\partial x_k^{in}} = a.
\]
Applying \eqref{dssout} and \eqref{dssin} one obtains
\[
2a\sum_i \Bigl[ A_{ik} - a \Bigl( x_i^{out} + x_k^{in} - \frac{1}{N} \Bigr) \Bigr] =0,
\]
and
\[
2a\sum_j \Bigl[ A_{kj} - a \Bigl( x_k^{out} + x_j^{in} - \frac{1}{N} \Bigr) \Bigr] =0,
\]
in which $\sum_i A_{ik} = k_k^{in}$ is the in-degree of the node $k$ and $\sum_j A_{kj} = k_k^{out}$ is its out-degree. Solving both equations for the properties $x_k^{out}$ and $x_k^{in}$ yields \[
x_k^{in} = \frac{k_k^{in}}{aN} \] and \[ x_k^{out} = \frac{k_k^{out}}{aN}.\]
Assuming the vectors of centralities $\mathbf{x}^{out}$ and $\mathbf{x}^{in}$ to have unitary 1-norm, i.e., $\sum_i x_i^{out} = \sum_i x_i^{in}=1$, one obtains $a=K_{tot}/N$ as in \eqref{a}, finally yielding
\begin{subequations}
\label{deg-dir}
	\begin{equation}
	\label{deg-in}
		x_k^{in} = \frac{k_k^{in}}{K_{tot}},
	\end{equation}
	\begin{equation}
	\label{deg-out}	
		x_k^{out} = \frac{k_k^{out}}{K_{tot}}.
	\end{equation} 
\end{subequations}
\eqref{deg-out}-\eqref{deg-in} correspond to rescaling the \textbf{out-degree} and \textbf{in-degree} by the total degree of the network.

\subsubsection{Unique contribution}
Let us start from the computation of the total unique contribution i.e., the $UC$ of the node $k$ when its properties \textit{out} and \textit{in} are considered together. From \eqref{f1dir}, one has
\begin{align*}
f(x_i^{out},0) &= ax_i^{out} - \frac{a}{N}; \\ \nonumber 
f(0, x_j^{in}) &= ax_j^{in} - \frac{a}{N}; \\ \nonumber 
f(0,0) &= - \frac{a}{N}. \nonumber 
\end{align*}
Using \eqref{uc-dir2}, one obtains
\begin{align*}
\Delta S\! S^{tot} = & \sum_{i \neq k} (-ax_k^{in})\Bigl( 2ax_i^{out} + ax_k^{in} -2\frac{a}{N} - 2A_{ik} \Bigr) \\ \nonumber 
& + \sum_{j \neq k} (-ax_k^{out}) \Bigl( 2ax_j^{in} +a x_k^{out} -2\frac{a}{N} - 2A_{kj} \Bigr) \\ \nonumber  &+ (-ax_k^{out} -ax_k^{in}) \Bigl( ax_k^{out} + ax_k^{in} -2\frac{a}{N} - 2 A_{kk} \Bigr) \\ \nonumber 
 =& -ax_k^{in} \sum_i \Bigl( -2ax_i^{out} - ax_k^{in} + 2\frac{a}{N} + 2A_{ik} \Bigr) \\ \nonumber & -ax_k^{out} \sum_j \Bigl(-2ax_j^{in} - ax_k^{out} + 2\frac{a}{N} + 2A_{kj} \Bigr) \\ \nonumber &+ 2a^2 x_k^{out} x_k^{in}, \nonumber
\end{align*}
in which the assumption $A_{kk}=0$ is used.
Substituting the values of $x_k^{out}$ and $x_k^{in}$ according to \eqref{deg-dir}, and considering $a=K_{tot}/N$, some algebra gives
\[
\Delta S\! S^{tot} = \frac{(k_k^{in})^2+(k_k^{out})^2}{N}+\frac{2 k_k^{in}k_k^{out}}{N^2} 
\]
from which the unique contribution is obtained
\begin{equation}
\label{uc-outin-deg}
UC_k^{tot} = \frac{1}{T\! S\! S} \Bigl[ \frac{(k_k^{in})^2+(k_k^{out})^2}{N}+\frac{2 k_k^{in}k_k^{out}}{N^2} \Bigr]
\end{equation}

The unique contribution obtained by separately considering the property \textit{out} or \textit{in} is found applying \eqref{dss-out} - \eqref{dss-in}, respectively. In this case one obtains
\begin{equation}
\label{uc-out-deg}
UC_k^{out} = \frac{(k_k^{out})^2}{N T\! S\! S},
\end{equation}
\begin{equation}
\label{uc-in-deg}
UC_k^{in} = \frac{(k_k^{in})^2}{N T\! S\! S}.
\end{equation}
Both the formulations in \eqref{uc-out-deg} and \eqref{uc-in-deg} are monotonic increasing function of $x_k^{out}$ and of $x_k^{in}$, respectively. Hence, ranking for increasing $UC_k^{out}$ and $UC_k^{in}$ values provide the same ranking as the classical in and out degree centrality.

\subsection{Hub-authority centrality}
\label{sec-hits}
Consider the estimator for directed network $f_2$ in Table 2, namely
\begin{equation}
\label{f2dir}
\hat{A}_{ik} =f_2(x_i^{out},x_k^{in}) =\gamma x_i^{out} x_k^{in}.
\end{equation}
Clearly,
\[
\frac{\partial f_2}{\partial x_k^{out}} = \gamma x_j^{out}, \hspace{1cm} \frac{\partial f_2}{\partial x_k^{in}} = \gamma x_i^{in}.
\]
Applying \eqref{dssout} and \eqref{dssin} one obtains
\[
\begin{cases}
\frac{\partial S\! S}{\partial x_k^{out}}= 2 \sum_j (\gamma A_{kj}x_j^{in} - \gamma^2 x_k^{out} (x_j^{in})^2)=0, \vspace{0.2cm} \\ 
\frac{\partial S\! S}{\partial x_k^{in}}= 2 \sum_i (\gamma A_{ik} x_i^{out} - \gamma^2 (x_i^{out})^2 x_k^{in})=0.
\end{cases}
\]
that, solved with respect to the properties $x_k^{out}$ and $x_k^{in}$, within the assumption of unitary 2-norm of the vectors, i.e. $\sum_i (x_i^{out})^2=1$ and $\sum_j (x_j^{in})^2=1$, yields
\begin{equation}
	\label{iteroutin}
	\begin{cases}
		x_k^{out} = \frac{1}{\gamma}\sum_j A_{kj}x_j^{in},\vspace{0.2cm} \\
		x_k^{in} = \frac{1}{\gamma}\sum_i A_{ik}x_i^{out}.
	\end{cases}
\end{equation}
In matrix form, 
\[
\begin{cases}
\gamma \mathbf{x}^{out} = \mathbf{Ax}^{in},\vspace{0.2cm} \\
\gamma \mathbf{x}^{in} = \mathbf{A^T}\mathbf{x}^{out}.
\end{cases}
\]
Some algebra gives
\[
\begin{cases}
\gamma^2 \mathbf{x}^{out} = \mathbf{AA^T}\mathbf{x}^{out},\vspace{0.2cm} \\
\gamma^2 \mathbf{x}^{in} = \mathbf{A^T A}\mathbf{x}^{in}.
\end{cases}
\]
Introducing the matrices $\mathbf{C}=\mathbf{A^T}\mathbf{A}$ and $\mathbf{D}=\mathbf{A}\mathbf{A^T}$, one has
\begin{subequations}
\label{xhits}
	\begin{equation}
	\label{xout}
		\gamma^2 \mathbf{x}^{out} = \mathbf{D}\mathbf{x}^{out},
	\end{equation}
	\begin{equation}
	\label{xin}	
		\gamma^2 \mathbf{x}^{in} = \mathbf{C}\mathbf{x}^{in}.
	\end{equation} 
\end{subequations}
\eqref{xhits} states that $\mathbf{x}^{out}$ and $\mathbf{x}^{in}$ are the dominant eigenvectors of the matrices $\mathbf{D}$ and $\mathbf{C}$, 
respectively, associated to the principal eigenvalue of the two matrices, such that $\gamma^2 = \lambda_1(\mathbf{C}) = \lambda_1(\mathbf{D}) = \sigma_1^2(\mathbf{A})$ \cite{kaplan1970,golub2012}, being $\sigma_1$ the principal singular value of the matrix $\mathbf{A}$. The formulation in \eqref{xhits} matches the \textbf{HITS algorithm} \cite{kleinberg1999}, used to identify \textit{hubs} and \textit{authorities} in networks. 

\subsubsection{Unique contribution}
First, consider the unique contribution to be computed with respect to both the properties. Using \eqref{f2dir}, one has
\[
f(x_i^{out},0) = f(0,x_j^{in}) = f(0,0) = 0,
\]
from which \eqref{uc-dir1} becomes
\begin{align*}
\Delta S\! S^{tot} =& \sum_{i \neq k} \Bigl[ -(\gamma x_i^{out}x_k^{in})^2 + 2 \gamma x_i^{out}x_k^{in}A_{ik} \Bigr] \\ \nonumber  &+ \sum_{j \neq k} \Bigl[ -(\gamma x_k^{out}x_j^{in})^2 + 2 \gamma x_k^{out}x_j^{in}A_{kj} \Bigr] \\ \nonumber &+ \Bigl[ -(\gamma x_k^{out}x_k^{in})^2 + 2 \gamma x_k^{out}x_k^{in}A_{kk} \Bigr] \\ \nonumber = & \sum_i \Bigl[ -(\gamma x_i^{out}x_k^{in})^2 + 2 \gamma x_i^{out}x_k^{in}A_{ik} \Bigr] \\ \nonumber &+ \sum_j \Bigl[ -(\gamma x_k^{out}x_j^{in})^2 + 2 \gamma x_k^{out}x_j^{in}A_{kj} \Bigr] \\ \nonumber &- \Bigl[ -(\gamma x_k^{out}x_k^{in})^2 \Bigr], \nonumber
\end{align*}
in which the assumption $A_{kk}=0$ is used. Some algebra provides
\begin{align}
\label{eq2-22}
\Delta S\! S^{tot} =& -\gamma (x_k^{in})^2 \sum_i (x_i^{out})^2 + 2 \gamma x_k^{in}\sum_i x_i^{out}A_{ik} \\ \nonumber & -\gamma (x_k^{out})^2 \sum_j (x_j^{in})^2 + 2 \gamma x_k^{out}\sum_j A_{kj}x_j^{in} \\ & + (\gamma x_k^{out}x_k^{in})^2. \nonumber 
\end{align}
Since the 2-norm of the vectors $\mathbf{x}^{out}$ and $\mathbf{x}^{in}$ is unitary and using \eqref{iteroutin}, one has
\[
\Delta S\! S^{tot} = \gamma^2 (x_k^{out})^2 + \gamma^2 (x_k^{in})^2 + (\gamma x_k^{out} x_k^{in})^2.
\]
The total unique contribution of the node $k$ applying the definition \eqref{UC} is
\begin{equation}
\label{uchits}
UC_k^{tot} = \frac{\gamma^2 (x_k^{out})^2 +\gamma^2(x_k^{in})^2+(\gamma x_k^{out}x_k^{in})^2}{T\! S\! S}.
\end{equation}

In order to compute the unique contribution accounting separately for the properties \textit{out} or \textit{in},  \eqref{dss-out} - \eqref{dss-in} are used
\[
\Delta S\! S^{out} = \sum_j \Bigl[ -(\gamma x_k^{out}x_j^{in})^2 + 2 \gamma x_k^{out}x_j^{in}A_{kj} \Bigr],
\]
and
\[
\Delta S\! S^{in} = \sum_i \Bigl[ -(\gamma x_i^{out}x_k^{in})^2 + 2 \gamma x_i^{out}x_k^{in}A_{ik} \Bigr].
\]
Going through the same algebra as for \eqref{eq2-22} and applying the definition of unique contribution, one obtains
\begin{equation}
\label{uc-auth}
UC_k^{out} = \frac{\gamma^2 (x_k^{out})^2}{T\! S\! S}.
\end{equation}
and
\begin{equation}
\label{uc-hub}
UC_k^{in} = \frac{\gamma^2(x_k^{in})^2}{T\! S\! S}.
\end{equation}
Both the formulations in \eqref{uc-auth} and \eqref{uc-hub} are monotonic increasing function of $x_k^{out}$ and of $x_k^{in}$, respectively. Hence, ranking for increasing $UC_k^{out}$ and $UC_k^{in}$ values provide the same ranking as the classical hub-authority algorithm.

\subsection{Multi-component centrality}
In the case of directed networks, the multi-component estimator is a function of the $s$-dimensional vectors $\mathbf{x}_i^{out}$ and $\mathbf{x}_j^{in}$ considered for evaluating node's importance, namely $\hat{A}_{ij} =f(\mathbf{x}_i^{out},\mathbf{x}_j^{in})$, where $\mathbf{x}_i^{out} = [x_{i,1}^{out},...,x_{i,s}^{out}]$ and $\mathbf{x}_j^{in} = [x_{j,1}^{in},...,x_{i,s}^{in}]$ . Within this framework, the multidimensional estimator is
\begin{align}
\label{f_seq_dir}
\hat{A}_{ij}(s) &= \gamma_1 x_{i,1}^{out}x_{j,1}^{in} +\gamma_2 x_{i,2}^{out}x_{j,2}^{in}+...+ \gamma_s x_{i,s}^{out}x_{j,s}^{in} \\ \nonumber 
&= \sum_{t=1}^s \gamma_s x_{i,t}^{out} x_{j,t}^{in}.
\end{align}

We assume the 2-norm of each vector $\mathbf{x}_{t}^{out}=[x_{1,t}^{out},...,x_{N,t}^{out}]$ and $\mathbf{x}_{i,t}^{in}=[x_{1,t}^{in},...,x_{N,t}^{in}]$ is unitary i.e., $\sum_i (x_{i,t}^{out})^2=\sum_i (x_{i,t}^{in})^2=1$. Moreover, we set an orthogonality condition between any two vectors $\mathbf{x}_t^{out/in}$ and $\mathbf{x}_{t^*}^{out/in}$, i.e.
\begin{equation}
\label{orthout}
\sum_i x_{i,t}^{out} \cdot x_{i,t^*}^{out} = 0, \hspace{1cm} \forall t \neq t^*,
\end{equation}
\begin{equation}
\label{orthoin}
\sum_i x_{i,t}^{in} \cdot x_{i,t^*}^{in} = 0, \hspace{1cm} \forall t \neq t^*.
\end{equation}

Similarly to Sect. \ref{sect-1}, in this multi-component setting the function $S\! S$ is expressed as \eqref{sskt}. In order to compute the centrality values, it is necessary to derive the function $S\! S_{k,t}$ accounting for the $s$ dimensions embedded in the estimators. The derivatives of the multi-component estimator \eqref{f_seq_dir} with respect to the variables $\mathbf{x}_{k,t^*}^{out}$ and $\mathbf{x}_{k,t^*}^{in}$ at any order $t^*$ are
\[
\frac{\partial f(\mathbf{x}_{i}^{out},\mathbf{x}_{k}^{in})}{\partial x_{k,t^*}^{in}} = \gamma_{t^*}x_{i,t^*}^{out},
\]
and
\[
\frac{\partial f(\mathbf{x}_{k}^{out},\mathbf{x}_{j}^{in})}{\partial x_{k,t^*}^{out}} = \gamma_{t^*}x_{j,t^*}^{in},
\]
that, introduced in \eqref{dssout} and \eqref{dssin}, provide
\begin{align*}
&2\sum_i \Bigl[ A_{ik} - \sum_t \gamma_t x_{i,t}^{out}x_{k,t}^{in} \Bigr]\gamma_{t^*}x_{i,t^*}^{out} = \\ \nonumber 
& \sum_i A_{ik} x_{i,t^*}^{out} - \sum_t \gamma_t x_{k,t}^{in} \sum_i x_{i,t^*}^{out}x_{i,t}^{out} =0 \nonumber 
\end{align*}
and
\begin{align*}
&2\sum_j \Bigl[ A_{kj} - \sum_t \gamma_t x_{k,t}^{out}x_{j,t}^{in} \Bigr]\gamma_{t^*}x_{j,t^*}^{in} = \\ \nonumber 
& \sum_j A_{kj} x_{j,t^*}^{in} - \sum_t \gamma_t x_{k,t}^{out} \sum_j x_{j,t^*}^{in}x_{j,t}^{in} =0 \nonumber 
\end{align*}
Using the conditions of othonormality, \eqref{orthout} - \eqref{orthoin}, some algebra provides
\begin{equation}
\label{svd}
\begin{cases}
		x_{k,t}^{out} = \frac{1}{\gamma_t}\sum_j A_{kj}x_{j,t}^{in},\vspace{0.2cm} \\
		x_{k,t}^{in} = \frac{1}{\gamma_t}\sum_i A_{ik}x_{i,t}^{out}.
	\end{cases}
\end{equation}
\eqref{svd} states that at any order $t$, the vectors $\mathbf{x}_t^{out}=[x_{1,t}^{out},...,x_{N,t}^{out}]$ and $\mathbf{x}_t^{in}=[x_{1,t}^{in},...,x_{N,t}^{in}]$ are the left and right singular vectors associated to the singular value $\gamma_t$, respectively. 

The estimation provided in \eqref{f_seq_dir} is the \textit{s-order low-rank approximation} of the original adjacency matrix $\hat{A}$. 

\subsubsection{Unique contribution}
In the multi-component setting for directed networks, the unique contribution is found accounting for the $s$ dimensions embedded in the estimator function $f$ (see \eqref{f_seq_dir}). In this case, when excluding the generic node $k$ from the estimation, all the properties $x_{k,t}^{out}$ and $x_{k,t}^{in}$, with $t=(1,...,s)$, are nullified. This yields
\[
f(\mathbf{x}_i^{out},0) = f(0,\mathbf{x}_j^{in}) =f(0,0) =0.
\]
Within this multi-component setting, the unique contribution can be computed with respect to both the properties $\mathbf{x}_{k,t}^{out}$ and $\mathbf{x}_{k,t}^{in}$, or with respect to one of the two. 

If both the properties are considered, \eqref{uc-dir2} holds, providing
\begin{align*}
\Delta S\! S^{tot} =& \sum_{i \neq k} \Bigl( - \sum_{t=1}^s \gamma_t x_{i,t}^{out} x_{k,t}^{in} \Bigr)\Bigl(\sum_{t=1}^s \gamma_t x_{i,t}^{out} x_{k,t}^{in}-2A_{ik} \Bigr) \\ \nonumber 
& + \sum_{j \neq k} \Bigl( -\sum_{t=1}^s \gamma_t x_{k,t}^{out} x_{j,t}^{in} \Bigr)\Bigl(\sum_{t=1}^s \gamma_t x_{k,t}^{out} x_{j,t}^{in} -2A_{kj} \Bigr) \\ \nonumber 
& + \Bigl(-\sum_{t=1}^s \gamma_t x_{k,t}^{out} x_{k,t}^{in} \Bigr)\Bigl( \sum_{t=1}^s \gamma_t x_{k,t}^{out} x_{k,t}^{in} -2A_{kk} \Bigr). \nonumber
\end{align*}
that is equivalent to
\begin{align*}
\Delta S\! S^{tot} =& \sum_i \Bigl[- \Bigl(\sum_{t=1}^s \gamma_t x_{i,t}^{out} x_{k,t}^{in} \Bigr)^2 +2A_{ik} \sum_{t=1}^s \gamma_t x_{i,t}^{out} x_{k,t}^{in} \Bigr] \\ \nonumber 
& + \sum_j \Bigl[ -\Bigl(\sum_{t=1}^s \gamma_t x_{k,t}^{out} x_{j,t}^{in} \Bigr)^2 -2A_{kj}\sum_{t=1}^s \gamma_t x_{k,t}^{out} x_{j,t}^{in} \Bigr] \\ \nonumber 
& + \Bigl(\sum_{t=1}^s \gamma_t x_{k,t}^{out} x_{k,t}^{in} \Bigr)^2. \nonumber
\end{align*}
Some algebra provides
\begin{align}
\label{eq3-25}
\Delta S\! S^{tot} =& -\sum_{t=1}^s \gamma_t (x_{k,t}^{in})^2 \sum_i (x_{i,t}^{out})^2 + 2 \sum_{t=1}^s \gamma_t x_{k,t}^{in} \sum_i A_{ik} x_{i,t}^{out} \\ \nonumber 
& - \sum_{t=1}^s \gamma_t (x_{k,t}^{out})^2 \sum_j (x_{j,t}^{in})^2 + 2 \sum_{t=1}^s \gamma_t x_{k,t}^{out} \sum_j A_{kj} x_{j,t}^{in} \\ \nonumber 
& + \Big( \sum_{t=1}^s \gamma_t x_{k,t}^{out} x_{k,t}^{in}\Bigr)^2. \nonumber
\end{align}
Using the orthonormality conditions \eqref{orthout} - \eqref{orthoin} and the formulation in \eqref{svd}, the unique contribution in the case of the multi-component estimator in directed networks is obtained
\begin{equation}
\label{uc-mult-dir}
UC(s)_k^{tot} = \sum_{t=1}^s \gamma_t^2 \Bigl( (x_{k,t}^{in})^2 + (x_{k,t}^{out})^2 \Bigr) + \Bigl( \sum_{t=1}^s \gamma_t x_{k,t}^{out} x_{k,t}^{in}\Bigr)^2.
\end{equation}

The unique contribution when accounting separately for the \textit{out} and \textit{in} properties, applying \eqref{dss-out} and \eqref{dss-in}, reads
\[
\Delta S\! S^{out} = \sum_j \Bigl[ -\Bigl(\sum_{t=1}^s \gamma_t x_{k,t}^{out} x_{j,t}^{in} \Bigr)^2 -2A_{kj}\sum_{t=1}^s \gamma_t x_{k,t}^{out} x_{j,t}^{in} \Bigr]
\]
and
\[
\Delta S\! S^{in} =  \sum_i \Bigl[- \Bigl(\sum_{t=1}^s \gamma_t x_{i,t}^{out} x_{k,t}^{in} \Bigr)^2 +2A_{ik} \sum_{t=1}^s \gamma_t x_{i,t}^{out} x_{k,t}^{in} \Bigr].
\]
Going through some algebra and applying the definition in \eqref{UC}, one has
\begin{equation}
\label{uc-out-md}
UC(s)_k^{out}= \frac{1}{T\! S\! S}\sum_{t=1}^s \gamma_t^2 (x_{k,t}^{out})^2,
\end{equation}
and
\begin{equation}
\label{uc-in-md}
UC(s)_k^{in}= \frac{1}{T\! S\! S}\sum_{t=1}^s \gamma_t^2 (x_{k,t}^{in})^2.
\end{equation}

\subsection{Estimation results}
\label{res_dirette}
\begin{center}
\begin{figure}[t]
\includegraphics[width=8cm]{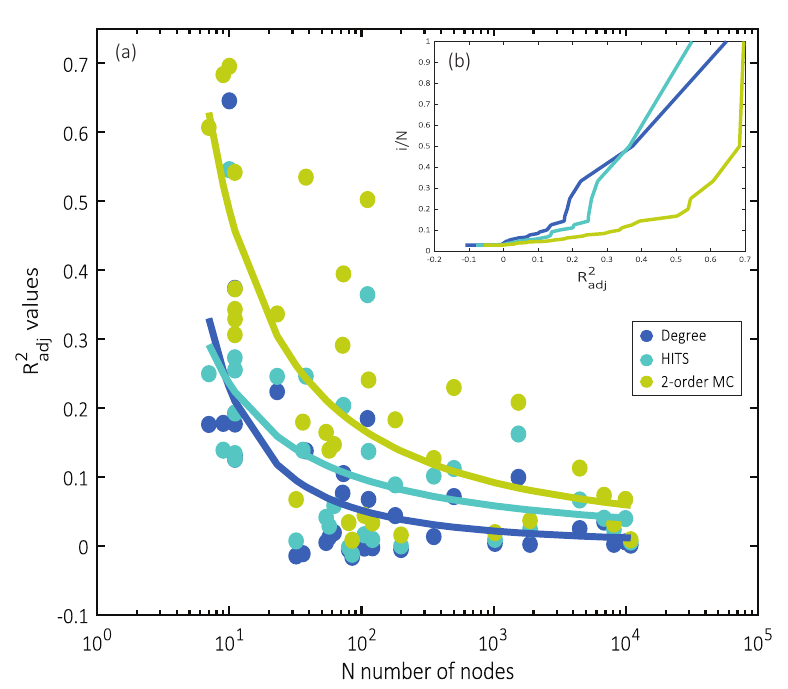}
\caption{\textbf{(a)} Values of the coefficient of determination $R^2_{adj}$ in semi-log scale obtained through the centrality-based estimators degree, hub-authority and multi-component (MC). Each dot refer to a directed network in the \textit{Sparse Matrix} database \cite{davis2011}. Power-law curves are fitted to the data to facilitate visual comparison. \textbf{(b)} Cumulative frequency curves for the $R^2_{adj}$ obtained by the three estimators.}\label{fig:rr_dir}
\end{figure}
\end{center}

We tested our framework on 36 networks freely available on the \textit{Suite Sparse Matrix Collection} \cite{davis2011}. The results obtained from our tests are shown in Fig.\ref{fig:rr_dir}.

The values of adjusted coefficient of determination, $R^2_{adj}$ are higher than those shown in Fig. 1, which were obtained from the application of our framework to undirected networks. This is mainly due to the fact that we are using two properties to characterize each node. As a consequence, the estimators (see Table 2) applied in case of directed networks project the information of the adjacency matrix from $N^2$ to $2N$, reducing the information gap. Also for directed networks, the one-component estimators perform poorly with respect to the two-component estimator. The hub-authority algorithm, however, has better performances than the degree, in particular when considering larger networks.

\phantomsection
\section*{Acknowledgments} The authors acknowledge ERC funding from the \textit{CWASI} project (ERC-2014-CoG, project 647473).

\phantomsection
\bibliographystyle{unsrt}
\bibliography{sample}


\end{document}


\flushbottom 

\maketitle 


\thispagestyle{empty} 


\section*{Introduction} 

Centrality aims to quantify the importance of nodes in a network \cite{Newman}. A first definition of this property dates back to the 50's, when it was introduced to study the role of nodes in communication patterns \cite{bavelas,leavitt1951}. During the following years, progress in social science provided several algorithms to evaluate nodes' centrality. These methods were tipically obtained through case-specific considerations about the functioning of social networks, mainly based on reasonings about how information spreads across people in a group \cite{bavelas}, and afterwards they were extended to other networks. Examples include the degree centrality \cite{shaw,freeman}, the Katz centrality \cite{katz1953}, the eigenvector centrality \cite{Bonacich1972}, the betweeness \cite{freeman,newman2005} and the closeness centrality \cite{freeman}, the PageRank \cite{pagerank}, the subgraph centrality \cite{estrada2005}, and the total communicability \cite{benzi2013}. Each metric defines node's centrality on the basis of some topological features of the considered node, such as the number of its connections, the connections of its neighbors, the number of walks and paths going across the node, etc. All the metrics hence provide different answers to the question \textit{``what does it mean to be central in a network ?”} (see, e.g., \cite{brandes2005,koschutzki2005,liao2017} for a literature review on centrality indexes and definitions). Due to the growing number of problems framed in network science, answering to the question about the meaning of node centrality is crucial for many scientific and technical field, ranging from epidemiology \cite{colizza2006,christakis2010,pastor2015} to economics \cite{guimera2005,schweitzer2009,hidalgo2009,tacchella2012}, from sociology \cite{borgatti2009} to engineering \cite{rinaldo2006,porta2009} and neuro-sciences \cite{bullmore2009,rubinov2010}.

Notwithstanding the need to have a measure of node relevance, an agreed definition of nodes' centrality is still lacking \cite{freeman}. The formulation of centrality metrics, in fact, typically descends from \textit{ad hoc} assumptions, where a node is said to be central if it has some specific feature which testifies its relevance in the network. For example, one may assume a node is more central if it has many connections with other nodes, which leads to the degree centrality as the natural measure. However, one may argue that nodes are not all equivalent, and that a weighted version of the degree of the nodes should be adopted, where the weight is the centrality itself: this leads to the eigenvector centrality as the adequate metric. Both these measures have a solid intuitive background. Nevertheless, one is left without the possibility of comparing the reliability of different measures of centrality, and therefore, of choosing which is the most effective metric -- and resulting node ranking -- for the specific problem at hand. 

Aiming at providing a more grounded deductive framework, we propose to tackle the centrality problem as a topology-estimation exercise. The proposed approach allows one (i) to deduce a hierarchy of metrics, (ii) to recast classical centrality measures (degree, eigenvector, Katz, hub-authority centrality) within a single theoretical scheme, (iii) to compare different centrality measures by evaluating their performances in terms of their capability to reproduce the network topology, and (iv) to extend the notion of centrality to a multi-component setting, still maintaining the possibility to use centrality to rank the nodes. 

This new perspective on centrality is general and can be applied to any network: undirected/directed, unweighted/weighted, and monopartite/bipartite networks. 

\section*{The new perspective: undirected, unweighted networks}
\label{undirected}

Let $G$ be an undirected, unweighted graph, with $N$ nodes and $E$ edges. $G$ is mathematically described by the symmetric adjacency matrix $\mathbf{A}$, whose $ij$-th element is $1$ if $i$ and $j$ share an edge, zero otherwise. Let $\hat{\mathbf{A}}$ be an estimator of the adjacency matrix. We expect a good estimator has larger $\hat{A}_{ij}$ values when $i$ and $j$ are connected (i.e., $A_{ij}=1$), and lower values otherwise (i.e., when $A_{ij}=0$). Our key idea is that the estimator of the generic element $A_{ij}$ should depend on some emerging property $x_i$ of the node $i$ and $x_j$ of the node $j$, with ($i,j$=1,...,$N$), representing the topological importance of each node, i.e. its centrality. In formulas, $\hat{A}_{ij}= f(x_i, x_j)$ where $f$ is an increasing function of both its arguments, since $\hat{A}_{ij}$ should increase when the nodes $i$ and $j$ are more ``central” in the network. Due to the symmetry of the matrix $\mathbf{A}$, the arguments of $f$ should also be exchangeable (i.e., $f(x_i,x_j)=f(x_j,x_i)$). Notice that the estimation process projects the information from $N^2$ to $N$ as we are estimating a $N \times N$ matrix using the $N$ values of nodes' centrality $x_i$. By definition, estimation is non exact, and $A_{ij} \neq \hat{A}_{ij}$. We suppose here that the error $\epsilon_{ij}$ related to the estimation is in additive form, namely
\begin{equation}
	\label{adderr}
		A_{ij}=\hat{A}_{ij}+\epsilon_{ij}=f(x_i,x_j)+\epsilon_{ij}.
\end{equation}

Under this perspective, the centrality measures can be obtained on sound statistical bases, as they arise as the result of a standard estimation problem. Different constraints about the error structure can be considered. The most classical approach -- least squares estimation -- entails minimizing the sum of the squared errors, i.e.
\begin{equation}
	\label{SS}
	S\! S (x_1,x_2,...,x_N) = \sum_i \sum_j \epsilon_{ij}^2 = \sum_i \sum_j (A_{ij} - f(x_i,x_j))^2.
\end{equation}
By minimizing this quantity with respect to $x_i$, i.e., solving the equation (see SI, Sect. 1)
\begin{equation}
\label{dss-und}
\frac{\partial S\! S}{\partial x_j} = 4 \sum_i \Bigr[ A_{ij}- f(x_i,x_j) \Bigr]~\frac{\partial f(x_i,x_j)}{\partial x_j}=0,
\end{equation}
a set of $N$ equations is obtained, which allows one to estimate the centrality value for all nodes
\footnote{The framework can be extended to consider the error term in \eqref{adderr} in multiplicative form, and/or to consider a node-wise unbiased constraint instead of minimizing $S\! S$.}. 

Within this statistical framework, the answer to the question \textit{``what does it mean to be central in a network ?”} is given through the analysis of the importance of the nodes in the estimation of $A_{ij}$: a node $i$ is more central than a node $j$ if the effect of its property $x_i$ on the minimization of $S\! S$ is larger i.e., if it is more ``useful” for estimating $\mathbf{A}$. Put it another way, the node $i$ is more important than the node $j$ if, when removing its property from the estimation of $A_{ij}$, the change in $S \! S$ recorded is higher than the one provoked by the exclusion of other nodes' property $x_j$. In order to account for this effect, we borrow the concept of the \textit{unique contribution} from the theory of commonality analysis  \cite{newton1967,nimon2010}. The unique contribution is a quantitative measure of the effect a single variable has in the estimation procedure \cite{nathans2012}. We define the unique contribution of the node $i$ as the gain in the coefficient of determination $R^2$ induced by considering $x_i$ in the estimation procedure. In formulas 
\begin{equation}
\label{UC}
	UC_i=R^2_{N} - R^2_{N-i} = \frac{S\! S_{N-i} - S\! S_{N}}{T\! S\! S},
\end{equation}
where $R^2 = 1 - \frac{S\! S}{T\! S \!S}$, with $S\! S$ as in \eqref{SS}, and $T\! S\! S= \sum_i \sum_j (A_{ij} - \bar{A})^2$, with $\bar{A}= 1/N^2 \sum_i \sum_j A_{ij}$ (see SI, Sect. 1.1 for details). The subscripts $N$ and $N-i$ in \eqref{UC} refer to the case when all the $x_i$ values are considered in the estimation (subscript $N$), or to the case when the $i$-th property is excluded (subscript $N-i$).

\begin{table*}[]
\setcellgapes{2pt}
\makegapedcells
\centering
\caption{Examples of the estimator functions $f$ to be set in \eqref{adderr} to obtain some commonly-used centrality measures. The unique contribution, which is here used to rank nodes for their centrality, is also reported. In the formulas, $K_{tot}=\sum_i \sum_j A_{ij}$ is the total degree of the network; $N$ is the number of nodes; $k_i=\sum_j A_{ij}$ is the degree of the node $i$; $\gamma$ and $B$ are two parameters whose values change according to the estimator function. In case of $f_2$, $\gamma$ equals the largest eigenvalue of $\mathbf{A}$. In case of $f_3$, $\gamma = 1/\alpha \sum_j x_j^2$ and $B =- 1/\sum_j x_j$, where $\alpha$ is the \textit{attenuation factor} of the Katz centrality. $T\! S\! S$ is defined in the text. Further details are given in SI, Sect.1.}
\label{tab1}
\begin{tabular}{cccc}
\toprule
\multicolumn{4}{c}{\textbf{Undirected networks}}                                                                                                                                                                                                                                                                                                                          \\ \midrule
\multicolumn{1}{c}{\textbf{Estimator function $f$}}                                                     & \multicolumn{1}{c}{\textbf{Centrality of node $i$}}                                                                       & \multicolumn{1}{c}{\textbf{Unique contribution of node $i$}}                                                & \textbf{Corresponding metric}  \\ \midrule
\multicolumn{1}{c}{$f_1 = \frac{K_{tot}}{N}~\left( x_i + x_j - \frac{1}{N}~\right)$} & \multicolumn{1}{c}{$x_i=\frac{k_i}{K_{tot}}$}                                                                     & \multicolumn{1}{c}{$UC_i=\frac{2(N+1)k_i^2}{N^2~T\! S\! S}$}                                                 & \textbf{Degree centrality}   \vspace{0.5cm}     \\ 
\multicolumn{1}{c}{$f_2 = \gamma x_i x_j$}                                           & \multicolumn{1}{c}{$x_i=\frac{1}{\gamma} \sum_j A_{ij}x_j$}                                                       & \multicolumn{1}{c}{$UC_i=\frac{\gamma x_i^2}{T\! S\! S}\left(\gamma x_i^2+2\gamma\right)$}                   & \textbf{Eigenvector centrality} \vspace{0.5cm}  \\ 
\multicolumn{1}{c}{$f_3 = \gamma x_i x_j + B$}                                       & \multicolumn{1}{c}{$x_i=\frac{\sum_j A_{ij}x_j}{\gamma \sum_j x_j^2} + \frac{B \sum_j x_j}{\gamma \sum_j x_j^2}$} & \multicolumn{1}{c}{$UC_i=\frac{\gamma x_i^2}{T\! S\! S}\left(\gamma x_i^2-2B+2\gamma \sum_j x_j^2 \right)$} & \textbf{Katz centrality}   \vspace{2pt}       \\ \bottomrule
\end{tabular}
\end{table*}

Different definitions of the function $f$ in \eqref{adderr} allow one to obtain different centrality metrics. Some noteworthy examples are described in Table \ref{tab1}. The degree centrality, the eigenvector centrality \cite{Bonacich1972} and the Katz centrality \cite{katz1953} are obtained by adapting very simple link-estimation functions. Recasting these centrality metrics into this new framework allows us to compare their performances, in terms of their ability to predict the adjacency matrix. New metrics can also be easily obtained, by adopting the estimator function $f$ which is the most suitable to represent the matrix-estimation problem at hand.

We would like to highlight that, even though the here proposed framework might remind of the networks models based on the fitness of the nodes \cite{caldarelli2002}, this is just a formal resemblance. In fact, while within the fitness model, the function $f$ represents a probability, in this framework it does represent an estimation of the $ij$-th element of the adjacency matrix.
\section*{Extending the new perspective}
\label{extensions}
A natural extension of the \textit{one-component} estimators (Table  \ref{tab1}) is to move toward more informative \textit{multi-component} metrics of nodes' centrality. 
The multi-component centrality considers more facets of the networks, by describing the role of network's nodes through more than one scalar property. In formulas $\mathit{\hat{A}_{ij}} = \mathit{f}(\mathbf{x}_i,\mathbf{x}_j)$, where $\mathbf{x}_i=[x_{i,1},...,x_{i,s}]$ is an $s$-dimensional vector embedding the $s$ properties of the node that should be considered for evaluating its importance (for $s=1$ the one-component metrics are recovered).

By taking the function $f_2$ in Table \ref{tab1} as the starting point for our reasoning \footnote{A multivariate extension of the function $f_1$ in Table \ref{tab1} is useless, because in the additive form the different components cannot bring independent information. An extension of $f_3$ would instead simply imply to add a constant value to \eqref{f_sequence}.}, a possible design of the multidimensional estimator is obtained,
\begin{equation}
\label{f_sequence}
\hat{A}_{ij}(s) = \gamma_1 x_{i,1}x_{j,1} +...+ \gamma_k x_{i,k}x_{j,k} +...+ \gamma_s x_{i,s}x_{j,s}.
\end{equation}
In this case, the estimation process projects $N^2$ data to $s \cdot N$, which is the number of independent variables used in the estimation. 
 
One may recognize that the formal structure of $\mathbf{\hat{A}}$ in \eqref{f_sequence} corresponds to the \textit{s-order low-rank approximation} of the matrix $\mathbf{A}$ \cite{golub2012}. Under a least squares constraint, one obtains that $\gamma_k$ is the $k$-th eigenvalue of the adjacency matrix and $\mathbf{x}_k=[x_{1,k},...,x_{N,k}]$ is its corresponding eigenvector (see SI, Sect. 1.5). 
Sorting the eigenvalues in descending order according to their absolute value,  eigenvectors of increasing order bring a monotonically decreasing amount of information. This solution corresponds to the \textit{Singular Value Decomposition} (SVD) \cite{golub2012} of the original matrix, truncated at the order $s$ (see SI, Sect. 1.5). The choice of the $s$ value therefore entails finding a good balance between the necessity to accurately describe the adjacency matrix and the willingness to have a parsimonious representation of a complex system. Different strategies can be pursued, also borrowing from the wide literature pertaining with the similar problem of deciding where to arrest the eigenvalue decomposition or the SVD (see, e.g., \cite{skillicorn2007} for a review). For example, one may choose the $s$ value corresponding to the first gap in the eigenspectrum of the adjacency matrix (see, e.g., \cite{iacobucci2017}). Alternatively, one may arrest the expansion in \eqref{f_sequence} when the explained variance reaches a predefined amount of the total variance of $\mathbf{A}$. This would entail that the remaining amount of variance is attributed to noise. 

The unique contribution of the $i$-th node, and hence its centrality value, when the expansion is arrested to $s$ is (see SI, Sect. 1.5.1)
\begin{equation}
\label{uc2d}
UC_i(s) = \frac{1}{T\! S\! S} \left[ \left(\sum_{k=1}^s \gamma_k x_{i,k}^2 \right)^2 + 2\sum_{k=1}^s \gamma_k^2 x_{i,k}^2 \right].
\end{equation}
It is clear that, by considering additional dimensions beyond the first, the node centrality ranking may significantly change, revealing node features which were hidden by the one-dimensional assumption. In fact, information on the structure and clustering of the network is contained in the eigenvectors beyond the first one (for more information see, e.g., \cite{newman2006, iacobucci2017}). In the case $s=N$, through the $UC$ one recovers the same ranking given by the degree centrality, since the approximated matrix equals the adjacency matrix, i.e. $\hat{\mathbf{A}}=\mathbf{A}$. It may be useful to note that the multi-component estimation of centrality, and the subsequent ranking given through the $UC$, entail a two-steps shrinkage of information. Firstly, the estimation projects data from $N^2$ to $s \cdot N$, and secondly the ranking projects from $s \cdot N$ to $N$. Therefore, the multi-component centrality acts as an additional pier for the bridge from $N^2$ to $N$, a pier which can be essential to pose the centrality estimation problem on more solid grounds. Clearly, both cases $s=1$ and $s=N$ correspond to limit situations when the additional pier is not in between $N^2$ and $N$, but it is on one of the two sides; in fact, in these situations one recovers the eigenvector centrality ($s=1$) and the degree centrality ($s=N$).
\section*{The new perspective: other network classes}

\subsection*{Directed, unweighted networks}
\label{directed}
In directed, unweighted networks, edges are directed and the elements $A_{ij}$ of the adjacency matrix $\mathbf{A}$ are $1$ if the edge points from $i$ to $j$, and zero otherwise. The adjacency matrix is generally asymmetric \cite{Newman}. 
\footnote{Notice that we here consider \textit{i} pointing to \textit{j}, i.e. the outgoing edges of the node \textit{i} are described onto the row \textit{i} of the matrix $\mathbf{A}$.}. In this kind of networks, nodes can be characterized by two properties, one concerning with the \textit{outgoing} centrality of the node, $x_i^{out}$, and the other concerning with the \textit{incoming} centrality, $x_i^{in}$. The estimator $\hat{A}_{ij}$ should depend on the outgoing centrality of node $i$ and on the incoming centrality of node $j$, namely $\hat{A}_{ij}=f(x_i^{out},x_j^{in})$. Examples of the \textit{out} and \textit{in} centrality of the nodes recovered in this statistical framework are the degree and the hub-authority centrality \cite{kleinberg1999} (see Table \ref{tab2}, details in SI, Sect. 2). Within this framework, the unique contribution can also be used to produce an overall ranking of network's nodes, combining both the \textit{out} and \textit{in} centrality of the nodes (see SI, Sect. 2). 

\begin{table*}[h]
\setcellgapes{2pt}
\makegapedcells
\centering
\caption{Estimator functions used for directed networks. In the formulas, $K_{tot}$ is the total degree of the network; $N$ is the number of nodes; $k_i^{out}$ and $k_i^{in}$ are the \textit{out} degree and \textit{in} degree of the node $i$; $\gamma$ is a parameter whose value equals the principal singular value $\sigma_1$ of $\mathbf{A}$. $T\! S\! S$ is defined in the text. The equations for the unique contribution are reported for the cases when outgoing and incoming properties of the node are separately considered (superscripts \textit{out} and \textit{in}), or for the case when they are considered together (superscript \textit{tot}). Further details are given in SI, Sect.2.}
\label{tab2}
\begin{tabular}{cccc}
\toprule
\multicolumn{4}{c}{\textbf{Directed networks}}                                                                                                                                                                                                                                                                                                                                                                                                                                         \\ \midrule
\multicolumn{1}{c}{\textbf{Estimator function $f$}}                                            & \multicolumn{1}{c}{\textbf{\begin{tabular}[c]{@{}c@{}}In and out\\ centrality of node $i$\end{tabular}}}                                                                                       & \multicolumn{1}{c}{\textbf{\begin{tabular}[c]{@{}c@{}}In and out\\ unique contribution of node $i$\end{tabular}}}                                                                                        & \textbf{\begin{tabular}[c]{@{}c@{}}Corresponding\\ metric\end{tabular}} \\ \midrule
\multicolumn{1}{c}{$f_1= \frac{K_{tot}}{N}~\left( x_i^{out} + x_j^{in} - \frac{1}{N}~\right)$} & \multicolumn{1}{c}{\begin{tabular}[c]{@{}c@{}}$x_i^{out} = \frac{k_i^{out}}{K_{tot}}$ \vspace{1cm} \\ $x_j^{in} = \frac{k_j^{in}}{K_{tot}}$ \end{tabular}}                   & \multicolumn{1}{c}{\begin{tabular}[c]{@{}c@{}}$UC_i^{out} = \frac{(k_i^{out})^2}{N~T\! S\! S},$ \hspace{0.3cm} $UC_i^{in} = \frac{(k_i^{in})^2}{N~T\! S\! S}$  \vspace{1cm} \\ $UC_i^{tot} = \frac{1}{T\! S\! S} \Bigl( \frac{(k_i^{out})^2+(k_i^{in})^2}{N}+\frac{2 k_i^{out} k_i^{in}}{N^2}\Bigr)$\end{tabular}}                                                                              & \textbf{\begin{tabular}[c]{@{}c@{}}Degree\\ centrality\end{tabular}} \vspace{1cm}\\ 
\multicolumn{1}{c}{$f_2 = \gamma x_i^{out} x_j^{in}$}                                          & \multicolumn{1}{c}{$\begin{cases}x_i^{out} = \frac{1}{\gamma}\sum_j A_{ij}x_j^{in} \vspace{0.5cm}\\x_j^{in} = \frac{1}{\gamma}\sum_i A_{ij}x_i^{out}\end{cases}$} & \multicolumn{1}{c}{\begin{tabular}[c]{@{}c@{}}$UC_i^{out} = \frac{(\gamma x_i^{out})^2}{T\! S\! S},$ \hspace{0.3cm} $UC_i^{in} = \frac{(\gamma x_i^{in})^2}{T\! S\! S}$  \vspace{1cm} \\ $UC_i^{tot} = \frac{1}{T\! S\! S} \Bigl[ \gamma^2 \Bigl( (x_i^{out})^2+(x_i^{in})^2 \Bigr)+ (\gamma x_i^{out} x_i^{in})^2 \Bigr]$\end{tabular}}                     & \textbf{\begin{tabular}[c]{@{}c@{}}Hub-authority\\ centrality \end{tabular}} \vspace{2pt} \\  \bottomrule
\end{tabular}
\end{table*}

The expansion to \textit{multi-component} centrality and estimator, is a function of the $s$-dimensional vectors of the nodes' properties $\mathbf{x}_i^{out}$ and $\mathbf{x}_j^{in}$, namely
\begin{equation}
\label{f_sdir}
\hat{A}_{ij}(s) =  \gamma_1 x_{i,1}^{out}x_{j,1}^{in} +...+ \gamma_k x_{i,k}^{out} x_{j,k}^{in} +...+ \gamma_s x_{i,s}^{out}x_{j,s}^{in}.
\end{equation}
\eqref{f_sdir} coincides with the Singular Value Decomposition (SVD) \cite{everett2013,golub2012}, being $\gamma_k$ the singular values and $\mathbf{x}_k^{out}$ and $\mathbf{x}_k^{in}$ the related singular vectors (see SI, Sect. 2.4). 

\subsection*{Weighted networks}
\label{weighted}
The extension of our approach to weighted networks is straightforward. It is in fact sufficient to replace in Eq. (1) - (3) the adjacency matrix $\mathbf{A}$ with the matrix of the weights $\mathbf{W}$ -- whose elements are defined as $w_{ij}>0$ if there is a flux connecting $i$ to $j$, zero otherwise -- and all the centrality measures in their weighted version are obtained as the solution of a matrix estimation exercise.

\subsection*{Bipartite networks}
\label{bipartite}
Bipartite networks are characterized by two sets of nodes - $\mathbf{U}$ and $\mathbf{V}$ - with $E$ edges connecting nodes between the two ensembles. These networks are described by the incidence matrix $\mathbf{B}$ whose elements $b_{ij}$ define the relationship between the nodes $i \in \mathbf{U}$ and the nodes $j \in \mathbf{V}$ \cite{Newman}. In this case, the estimator $\hat{B}_{ij}$ will be a function of a property $x_i$ of the nodes in the ensamble $\mathbf{U}$ and of a property $y_j$ of the nodes in the ensamble $\mathbf{V}$ i.e., $\hat{B}_{ij}=f(x_i,y_j)$. The centrality metrics obtained in Table \ref{tab2} are straightforward extended to bipartite networks. By using the function $f=\gamma x_i y_j$ and assuming a multiplicative error structure and an unbiased estimator, it is possible to recover the \textit{Fitness-Complexity} algorithm, extensively used in characterizing nations' wellness \cite{tacchella2012,eci}. Specifically, $x_i$ represents the Fitness of the node $i$ and $y_j$ the Complexity of node $j$.

\section*{Results and Discussion}
\label{application}
\begin{figure}
\centering
\includegraphics{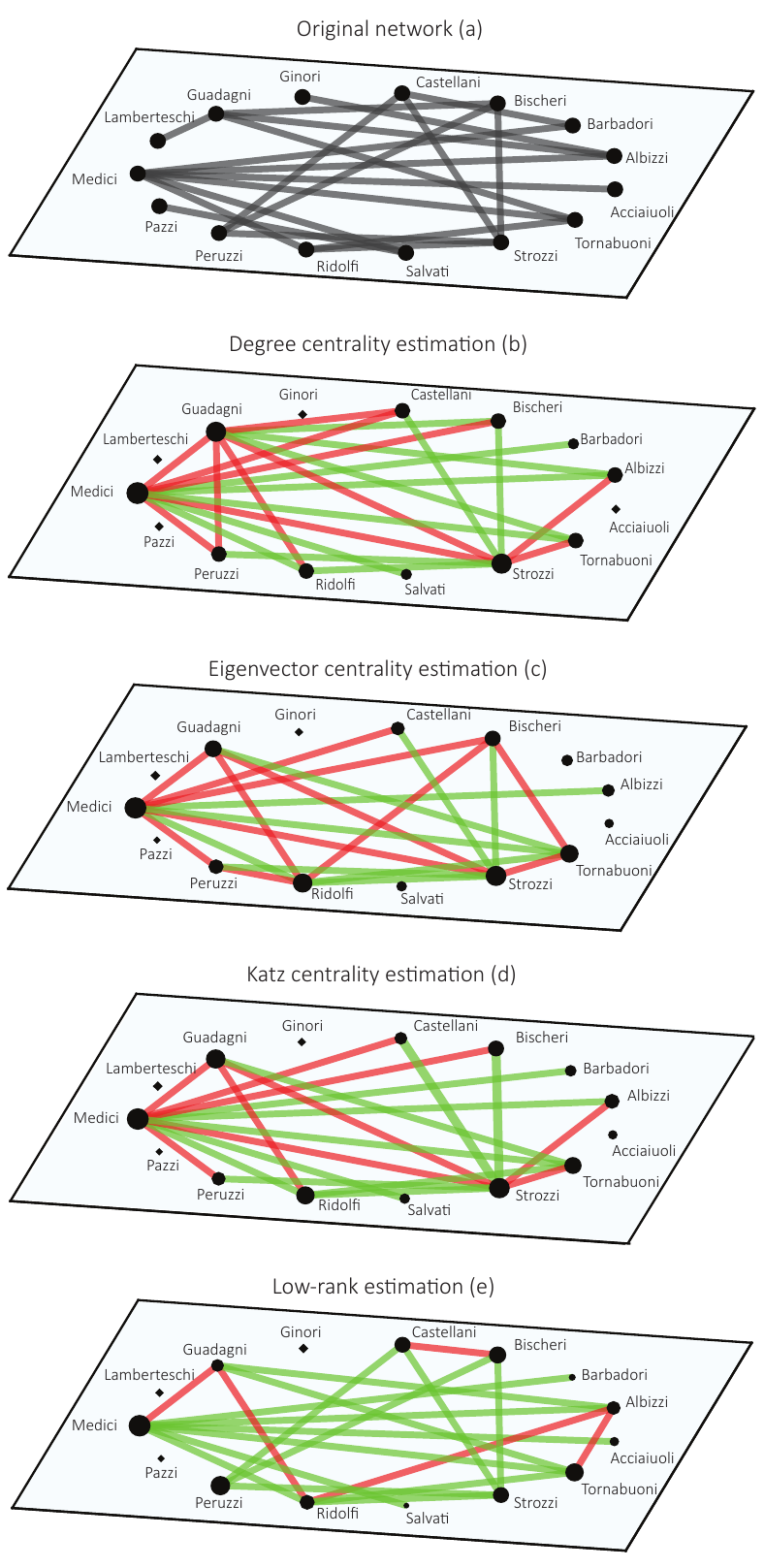}
\caption{Estimation results for the undirected network of Florentine Intermarriage Relations, represented in panel \textbf{(a)}. Panels \textbf{(b) - (d)} refer to the topology estimated by the degree, eigenvector, and Katz centrality, respectively. Panel \textbf{(e)} shows the estimated network as given by the multi-component estimator with two components ($s=2$). In the figure, correctly estimated links are highlighted in green, while spurious links are red colored. Nodes' size in panels \textbf{(b) - (e)} is proportional to the position in the ranking resulting from the unique contribution, ordering the list from least to most central node. We plot in Fig. \ref{fig:netestim} only the $E$ larger values of $\hat{A}_{ij}$, thus preserving in all the reconstructed networks the number $E$ of edges of the real network. Exception is made when the $E$-th larger value of $\mathbf{\hat{A}}$ is a tie, in which case more than $E$ edges are plotted.}\label{fig:netestim}
\end{figure}

We illustrate our new perspective starting in Fig. \ref{fig:netestim} with an analysis of the network of the Florentine Intermarriage Relations \cite{padgett1993}. The network has 15 nodes representing the most notables Renaissance families in Florence connected by marriage relations (20 edges). Within our framework, the centrality measures have a counterpart in a link-estimation function, which allows to perform a visual and numerical comparison with the original network. We plot the original network in Fig. \ref{fig:netestim}.\textbf{(a)}, and those resulting from the use of the one-component centrality measures in Fig. \ref{fig:netestim}.\textbf{(b-d)}. The centrality-based estimations are performed using the functions reported in Table \ref{tab1}. For the computation of the Katz centrality, we used $\alpha=0.5/\lambda_1$ following \cite{benzi2014}, being $\lambda_1$ the principal eigenvalue of $\mathbf{A}$ (see SI, Sect. 1.4). The network representation in Fig. \ref{fig:netestim}.\textbf{(e)} shows the result of the estimation provided by the multi-component estimation with $s=2$. Fig. \ref{fig:netestim} highlights the low agreement between the one-dimensional modeled networks and the real one. Several spurious and lacking links appear in the reconstructed graphs. The network representation is significantly improved when using the multi-component estimator ($s=2$) in Fig. \ref{fig:netestim}.\textbf{(e)}.

Besides the visual inspection, we compute the adjusted coefficient of determination $R^2_{adj}$ between the original and the estimated matrices, $\mathbf{A}$ and $\hat{\mathbf{A}}$, in order to measure the quality of the estimation. $R^2_{adj}$ is defined as
\[
R^2_{adj}=1-(1-R^2) \frac{N^2}{N^2-s\cdot N} = 1-(1-R^2)\frac{N}{N-s}.
\]
The choice of $R^2_{adj}$ as an error metric is consistent with the concept of unique contribution (see \eqref{UC}). Moreover, this error measure is applicable to binary variables as well and the ``adjusted” version of $R^2$ allows one to compare the results obtained from distinct estimators and on differently sized and structured networks. For the Florentine Intermarriage Relations network, the adjusted determination coefficient for the multi-component estimator is $R^2_{adj}=0.30$, while for the other estimators is around $R^2_{adj}=0.07$, confirming the outcomes of the visual inspection.

\begin{figure}
\includegraphics[width=1\linewidth]{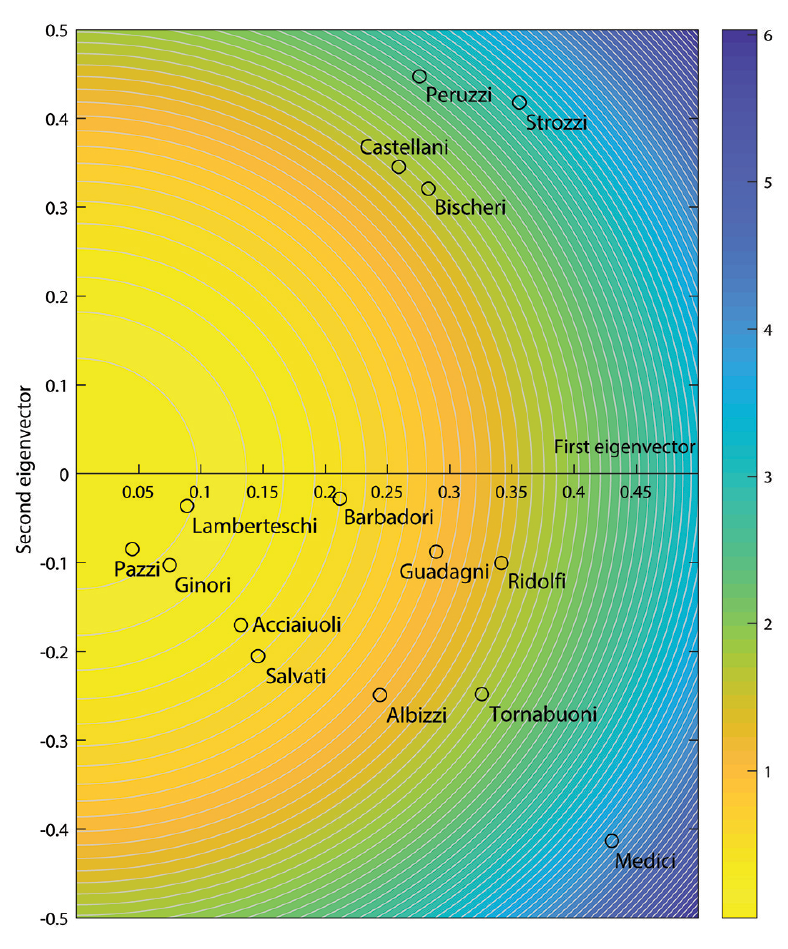}
\caption{Contour plot of the unique contribution resulting from the application of \eqref{uc2d} with $s=2$. The $x_{i,1}$ values (corresponding to the components of the first eigenvector) are on the x-axis, while the values of $x_{i,2}$ (related to the components of the eigenvector corresponding to the second eigenvalue, ordered following the method described in the SI, Sect. 1.5) are on the y-axis.  The open circles correspond to the $x_{i,1}$ and $x_{i,2}$ values for the Florentine Intermarriage Relations network.} \label{fig:tac}
\end{figure}

The three classical centrality metrics (degree, eigenvector, Katz) produce different rankings of the Florentine families. While the \textit{Medici} are always the top-ranked family, other families significantly change their position in the rankings (e.g., the ranking of the \textit{Ridolfi} family changes from $3$ to $8$ when different methods are considered). By embracing our new perspective on network centrality it is possible to compare these rankings claiming that, despite the differences, from a statistical point of view the three metrics bring the same information about the topology of the network. The need to extend the centrality concept toward multiple dimensions manifestly emerges from Fig. \ref{fig:tac}. The second eigenvector distinctly identifies the group constituted by the families \textit{Strozzi-Peruzzi-Castellani-Bischeri}, while highlighting how the \textit{Medici} family is left alone by these four families. In this case the information brought by the second eigenvector is clearly relevant in determining the ranking of the nodes. If one considers only the first eigenvector, \textit{Ridolfi} family would be ranked in the third position. More correctly, the addition of information carried by the second eigenvector, combined through the unique contribution, ranks the \textit{Ridolfi} in the seventh position. 

\begin{figure}
\includegraphics[width=1\linewidth]{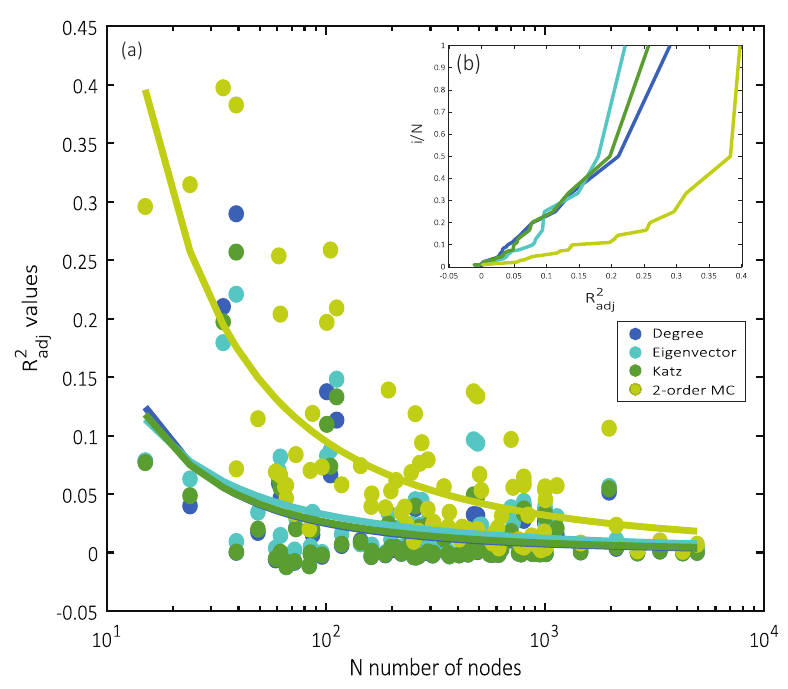}
\caption{\textbf{(a)} Values of the coefficient of determination $R^2_{adj}$, in semi-log scale obtained through the centrality-based estimators degree, eigenvector, Katz and multi-component (MC). Each dot refer to a network in the \textit{Sparse Matrix} database \cite{davis2011}. Power-law curves are fitted to the data to facilitate visual comparison. \textbf{(b)} Cumulative frequency curves for the $R^2_{adj}$ obtained by the four estimators.}\label{fig:rr}
\end{figure}

The outcomes of the analysis of the network of the Florentine Intermarriage Relations are fully confirmed by a more extended analysis on 106 undirected networks, all freely available at \url{https://sparse.tamu.edu/} \cite{davis2011}. The values of $R^2_{adj}$ obtained from the application of the functions in Table \ref{tab1} are reported in Fig. \ref{fig:rr}. Two features clearly emerge. Firstly, the degree, the eigenvector and the Katz centrality systematically perform poorly when considered under the perspective of estimating the networks topology. This is essentially due to the compression of information from $N^2$ to $N$ implied by the matrix-estimation exercise, undermining the performance of the estimators. In general, $R^2_{adj}$ decreases proportionally to the square root of $N$, following the behavior of the standard deviation of the centrality-based estimators. Hence, the largest the size, the more information is lost during the estimation. The plot shows systematically higher values of $R^2_{adj}$ resulting from the application of the two-components estimator \eqref{f_sequence}. As expected, considering more node's properties dramatically improves the estimation quality. Qualitatively similar results for directed networks are reported in the SI, Sect. 2.5.

A second key feature emerging from Fig. \ref{fig:rr} is that the values of $R^2_{adj}$ obtained from different one-component estimators are only slightly different from one another,  and there is no evidence of one centrality measure outperforming the others. It follows that, despite the different nature of the metrics (i.e., the degree is a \textit{local} measure of nodes' importance, while the eigenvector and the Katz centrality are \textit{global} measures \cite{liao2017}), all the metrics provide very similar and limited information about the topology of the networks. In this case, using different centrality metrics would not add new and divers information, resulting with redundancy of the metrics and therefore providing a further proof of their correlation \cite{schoch2017}.   

\section*{Conclusions}
\label{conclusions}
This work introduced a different point of view about centrality, through which the evaluation of the importance of nodes is recast as a statistical-estimation problem. Here, centrality becomes the node-property through which one estimates the adjacency matrix of the network, breaking new ground in the way we understand node centrality. Many of the most commonly used centrality metrics can be deduced within this theoretical framework, thus paving the way for an unprecedented chance to quantitatively compare the performances of different centrality measures. 

Aiming at showing the innovative power of our statistical perspective on centrality metrics, in this paper we focused on the application of this framework on monopartite networks. We stress that our approach is very general and should not be restricted to the examples reported above. Moreover, we argue that the estimator functions may also shed some light on the mathematical nature of the algorithms used to evaluate node centrality. In many cases, this would allow to find the exact analytic solution of the underlying mathematical maps and thus avoid tedious and imprecise iterative solutions.

Finally, the estimators could also explain the capability of the various algorithms to account for the nodes-nodes interactions. For example, by looking at the functions in Table \ref{tab1}, it is indeed clear that the degree centrality, obtained from a linear combination of the single nodes’ properties, cannot accommodate non-linear interactions among nodes. For this reason, the comparison of the performances of the various algorithms within our framework, could also be illuminating on the nature of the nodes interactions of a given system.

Tests on a large number of networks show that there are no outperforming one-dimensional, centrality-based estimators and that all the metrics provide poor information regarding networks' topology. Our results, within the context of the still ongoing debate on the centrality metrics and the associated ranking (in several fields, see, e.g.,  \cite{rothenberg1995,koschutzki2005,kiss2008,liao2017,pietronero2017}), provides a further proof that centrality metrics are highly correlated \cite{benzi2014,valente2008,perra2008,meghanathan2015,li2015} and that they provide similar information about the importance of the nodes. Within this new framework, a natural multi-component extension of node centrality emerges as a possible solution to improve the quality of the estimations and, subsequently, of node ranking. Our approach therefore provides a possible quantitative answer to the long-standing question \textit{``what does it mean to be central in a network ?”}. 

\phantomsection
\section*{Acknowledgments} The authors acknowledge ERC funding from the \textit{CWASI} project (ERC-2014-CoG, project 647473).

\phantomsection
\bibliographystyle{unsrt}
\bibliography{sample}
